\documentclass[conference,compsoc]{IEEEtran}
\ifCLASSOPTIONcompsoc
  \usepackage[nocompress]{cite}
\else
  \usepackage{cite}
\fi
\ifCLASSINFOpdf
\else
\fi
\usepackage{enumitem}
\usepackage{tikz}
\usepackage{tabularx}
\newcolumntype{Y}{>{\centering\arraybackslash}X}

\newcolumntype{Z}{>{\hsize=1.2\hsize}X}
\newcolumntype{Q}{>{\hsize=.8\hsize}X}
\newcolumntype{V}{>{\hsize=.15\hsize}X}
\usepackage{caption}
\usepackage{subcaption}
\usepackage{stfloats}
\usepackage{multirow}
\usepackage{multicol}
\usepackage{makecell}
\usepackage{balance}
\usepackage{hyperref}
\usepackage{nth}
\usepackage{xurl}
\usepackage{censor}
\usepackage{verbatim}
\usepackage{booktabs}
\usepackage{tikz}
\usetikzlibrary{shapes.geometric, arrows, positioning}

\tikzstyle{startstop} = [rectangle, rounded corners, minimum width=2.5cm, minimum height=1cm, text centered, draw=black, fill=red!30]
\tikzstyle{process} = [rectangle, minimum width=3cm, minimum height=1cm, text centered, draw=black, fill=orange!30]
\tikzstyle{note} = [rectangle, rounded corners, minimum width=3cm, minimum height=1cm, text centered, draw=black, fill=yellow!30]
\tikzstyle{data} = [rectangle, minimum width=2.5cm, minimum height=1cm, text centered, draw=black, fill=blue!30]
\tikzstyle{arrow} = [thick,->,>=stealth]

\begin{document}

% \title{Assessing Contextual Phone Scam Warnings in Naturalistic Settings: Insights from Blind/Low Vision and Sighted Individuals}

\title{(Blind) Users Really Do Heed Aural Telephone Scam Warnings}


 \makeatletter
    \newcommand{\linebreakand}{%
      \end{@IEEEauthorhalign}
      \hfill\mbox{}\par
      \mbox{}\hfill\begin{@IEEEauthorhalign}
    }
    \makeatother

\author{\IEEEauthorblockN{Filipo Sharevski}
\IEEEauthorblockA{School of Computing \\
DePaul University\\
Chicago, IL 60604\\
Email: fsharevs@depaul.edu}
\and
\IEEEauthorblockN{Jennifer Vander Loop}
\IEEEauthorblockA{School of Computing \\
DePaul University\\
Chicago, IL 60604\\
Email: jvande27@depaul.edu}
\linebreakand %
\IEEEauthorblockN{Bill Evans}
\IEEEauthorblockA{School of Computing \\
DePaul University\\
Chicago, IL 60604\\
Email: wevans9@depaul.edu}
\and
\IEEEauthorblockN{Alexander Ponticello}
\IEEEauthorblockA{CISPA Helmholtz Center for \\
Information Security\\
Germany\\
Email: alexander.ponticello@cispa.de}
}


% compsoc conferences compared to traditional conferences), use this
%Telephone: (800) 555--1212, Fax: (888) 555--1212}


\maketitle

\begin{abstract}
This paper reports on a study exploring how two groups of individuals, \textit{legally blind} (\textit{n}=36) and \textit{sighted} ones (\textit{n}=36), react to aural telephone scam warnings in naturalistic settings. As spoofing a CallerID is trivial, communicating the \textit{context} of an incoming call instead offers a better possibility to warn a receiver about a potential scam. Usually, such warnings are visual in nature and fail to cater to users with visual disabilities. To address this exclusion, we developed an \textit{aural} variant of telephone scam warnings and tested them in three conditions: baseline (no warning), short warning, and contextual warning that preceded the scam's content. We tested the two most common scam scenarios: \textit{fraud} (interest rate reduction) and \textit{identity theft} (social security number) by cold-calling participants and recording their action, and debriefing and obtaining consent afterward. Only two participants ``pressed one'' as the scam demanded, both from the \textit{legally blind} group that heard the contextual warning for the social security scenario. Upon close inspection, we learned that one of them did so because of accessibility issues with their screen reader and the other did so intentionally because the warning convinced them to waste the scammer's time, so they don't scam vulnerable people. Both the \textit{legally blind} and \textit{sighted} participants found the contextual warnings as powerful usable security cues that, together with STIR/SHAKEN indicators like \texttt{Scam Likely}, would provide robust protection against any type of scam. We also discussed the potential privacy implications of the contextual warnings and collected recommendations for usably accessible implementation. 

\end{abstract}


% \fi
\IEEEpeerreviewmaketitle

\section{Introduction} \label{sec:introduction}
Unwanted telephone calls inconvenience people on a daily basis because it's trivial for one to spoof the CallerID and to automate the process of calling a vast range of telephone numbers with relative ease~\cite{Du2023}. Most of these calls attempt to ``scam'' people, i.e., defraud them financially or steal their identifying information such as their Social Security number~\cite{Tu2016}. The Federal Trade Commission (FTC) shows the dire dimensions of this ``inconvenience'' as people in the US consistently incur a median loss of \$1200 and millions of them reported their identity stolen through various telephone scam calls on a yearly basis since 2020~\cite{ftc-data}.

Enabled through communication technologies, telephone scams are not entirely unlike emails or SMS text messages that attempt to socially engineer a target. In all cases, the adversary employs a persuasive pretext with the goal of eliciting an action from the target and attempts to maintain an impression of legitimacy. But telephone scams do differ in that they entail targets to decide to take a call or not ``on the spot'' and further, how to proceed. Protection-wise, thus, there is limited space for intervention. On an infrastructure level, an attempt has already been made to establish a call verification solution -- STIR/SHAKEN -- so a target call receiver could verify the authenticity of the CallerID ~\cite{Edwards2020}. Usually, the target call receiver is ``warned'' either through a visual indicator displayed on the call screen or the CallerID is labeled as \texttt{Scam Likely}. The solution, however, is not perfect and telephone call scammers find a way around it or the calls are often times incorrectly labeled~\cite{Pandit2023}.

On a device level, users could use call-blocking apps (e.g., Hiya) but this protection is inefficient as it demands a proactive list update of unwanted numbers. Recently, a new device-level protection was proposed by Google~\cite{Heather-Google-Scam} where the \textit{content} of an unwanted call is automatically screened using generative AI technology for known scam patterns. A visual warning -- akin to those Google assigns for suspicious emails~\cite{GooglePhishing} (see Figure~\ref{fig:visualwarning}) -- is then shown on the call screen to tell the receiver what the scam is about. Going beyond the call metadata and into the user call data, on a device level, is seen as necessary to effectively address the problem of spoofed CallerID and append the STIR/SHAKEN protection against scams. Aware of the privacy implication such a solution might have (more details in~\ref{sec:privacy-discussion})~\cite{Lomas-Google-Scam}, Google envisions this to be an \textit{opt-in} feature for future versions of their Android operating system~\cite{GoogleScamDetection}. 

Resembling Google's phishing or spam email banners, the telephone scam warning combines \textit{visual elements} (e.g., red color shades, ``End call'' button, etc.) and \textit{context} (e.g., ``\textit{Banks will never ask you to move money to keep it safe}'') to cue target call receivers away from the scam ``on the spot.'' While this intuitive approach might work for visually able people, past research shows that these warnings, as in the case of emails, are largely inaccessible for people with visual disabilities~\cite{Yu2023, Sharevski2024-usenix}. People with statutory (legal) blindness are equal targets of telephone scams as their sighted counterparts~\cite{sherman2021characterizing, Munyaka2024} so it is unknown whether contextual telephone scam warnings, such as Google's proposed solution, will have an accessible variant. 

Google's telephone scam avoidance is not available yet for any user or, for that matter, accessibility testing~\cite{GoogleScamDetection}. It did nonetheless inspire us, meanwhile, to independently develop \textit{aural} counterparts and test them in naturalistic settings (i.e., by cold-calling) with both legally blind and sighted telephone users in two variants (one \textit{short} and one long, \textit{contextual} warning) and two of the most prevalent real-world scamming scenarios (\textit{interest rate reduction} and \textit{Social Security} fraud). Based on the past evidence of visual call screen indicators' inaccessibility ~\cite{Munyaka2024, Sherman2020a11y, sherman2020you, sherman2021characterizing} we reasoned that is better to proactively develop and test aural options that are by default accessible and ready for use -- regardless of one's visual (dis)ability -- when (and if) the new way of combating telephone scams becomes widely available. To these objectives, we conducted an empirical study to answer the following research questions: 

\vspace{0.5em}

\begin{itemize}
    \itemsep 0.7em
    \item \textbf{RQ1:} How would \textit{legally blind} people initially respond to an (a) \textit{interest rate reduction} or (b) \textit{Social Security fraud} scam call, compared to their \textit{sighted} counterparts, in three conditions: (1) \textit{without} a warning (baseline); (2) with a \textit{short aural} warning; and (3) with a \textit{contextual aural} warning?

    % \item \textbf{RQ1b:} How would \textit{legally blind} people initially respond to a  \textit{social security} telephone scam call, compared to their \textit{sighted} counterparts,  in three conditions: (1) \textit{without} a warning (baseline); (2) with a \textit{short aural} warning; and (3) with a \textit{contextual aural} warning?

    % \item \textbf{RQ1c:} How would \textit{sighted} people initially respond to a telephone \textit{interest rate reduction} scam in three conditions: (1) \textit{without} a warning (baseline); (2) with a \textit{short aural} warning; and (3) with a \textit{contextual aural} warning?    

    % \item \textbf{RQ1d:} How would \textit{sighted} initially people respond to a telephone \textit{social security} scam call in three conditions: (1) \textit{without} a warning (baseline); (2) with a \textit{short aural} warning; and (3) with a \textit{contextual aural} warning?

    \item \textbf{RQ2:} How do (a) \textit{legally blind} and (b) \textit{sighted} people usually detect and deal with telephone scam calls?

    \item \textbf{RQ3:} What usability preferences do (a) \textit{legally blind} and (b) \textit{sighted} people have about telephone scam warnings?

    \item \textbf{RQ4:} What privacy concerns and design recommendations do (a) \textit{legally blind} and (b) \textit{sighted} people have about telephone scam aural warnings?

\end{itemize}

\vspace{0.5em}

The key takeaways from our study are that the \textit{aural} warnings in both short and contextual variants: (1) work, as only two people in the \textit{legally blind} group ``pressed one'' and only did so either because of a problem with the screen reader or intentionally to waste the scammers time; (2) combined with the area code of the CallerID, offer a robust cuing mechanism against telephone scam calls independently from the scam's pretext and fraud type, for example, financial or identity theft). The STIR/SHAKEN indicator (e.g., \texttt{Scam Likely}) does help fend off scam calls, but we found that it creates a problem where the screen readers don't verbalize the actual CallerID in cases when the call is wrongfully labeled, leaving \textit{legally blind} users without the opportunity to accept a legitimate call. Privacy equally mattered for our participants, with the lack of control and transparency being the topmost concerns. Design-wise, the aural warnings were well received, with few recommendations for haptic adjustments and text comprehensibility. 

\noindent \textbf{The main contributions of this paper are}: 

\begin{itemize}
\itemsep 0.5em
    \item Empirical evidence of the way \textit{legally blind} individuals detect, screen, and respond to telephone scams -- in naturalistic settings -- involving aural telephone warnings compared to their \textit{sighted} counterparts;  

    \item A novel, usably accessible telephone scam protection approach that cues receivers about the \textit{context} or the scam's intent (fraud or identity theft); 

    \item A set of privacy concerns as barriers for inferring the \textit{context} of a scam call through AI means and associated design recommendations that consider both the accessibility needs and usability habits of individuals target of telephone scams.

\end{itemize}

% The \textit{legally blind} individuals, thus, far more relied on \textit{in-call} cues, such synthesized voice, delay in speech, and the content of the scam itself. The participants in this group had a higher preference for the contextual warning in the \textit{social security} scam scenario. The \textit{sighted} participants, on the other hand, preferred the contextual variant in both scenarios, though most of them ultimately noted would opt for a visual contextual alert on their home screen. 

\section{Background} \label{sec:background}
\subsection{Telephone Scams}
Receiving unsolicited calls with pre-recorded messages became an unavoidable burden for the vast majority of people in the US, despite efforts to curb their placement on both the infrastructural, regulatory, and user side~\cite{Tu2016}. Some of these calls are automated -- colloquially referred to as ``robocalls'' -- though not all of them are unsolicited or unwanted. For example, people regularly receive robocalls about their doctor appointment reminders, prescriptions, or political campaigns~\cite{Edwards2020}. Some of these calls are spam such as telemarketing, sales calls, or polls~\cite{ftc-ndncr}. Some of these calls are \textit{scams}, or unwanted ones aiming to exhort money or compromise the recipient's private information.

Telephone scams, usually coming from a spoofed caller ID, pose a unique problem because they could be automated and placed as ``robocalls'' and/or could hide behind a seemingly innocuous spam pretext. Targeting the spoofing ID part, call authentication solutions such as STIR/SHAKEN have been introduced to help users ``verify'' numbers to assert a call is coming from a source it says it is~\cite{Edwards2020}. To help with the automation and spamming parts, the FTC appended the curbing effort by introducing the national ``Do Not Call Registry'' list where users could enter their number to be spared from unsolicited calls~\cite{ftc-ndncr}. Phone companies, device manufacturers, and third-party application developers also offered call-blocking and call-labeling services, so users can receive as few unwanted calls as possible~\cite{ftc-block}. 

But even with such extensive and layered protection, telephone scams still reach users in alarming numbers. Just in the first half of 2024, the FTC received around half a million complaints about telephone scams that successfully scammed one in five users for a staggering \$229 million in losses and more than 132,000 identities stolen~\cite{ftc-data}. Fundamentally, the problem with telephone scams is hard to solve due to a couple of factors: (i) the nature of phone calls as a communication service; and (ii) the protection structure itself. Telephone scams in essence are unwanted communication requests not unlike unsolicited emails or SMS texts (spam, phishing, and such). While (scam) emails and SMS texts are \textit{non-real} time in nature --  the recipient need not to respond promptly -- the telephone (scam) calls require the recipient to pick up the phone and respond in \textit{real} time. One could ``screen'' the call and let it go to a voice mail, but the aforementioned FTC numbers suggest that that's not a common practice (besides, scammers tend to ``pressure'' recipients with repeated voicemail spam in order for them to pick up the phone~\cite{Prasad2020}). 

In other words, a recipient of a telephone scam must decide ``on the spot'' whether the call is unwanted or unsolicited in the first place and take action to terminate or proceed~\cite{Tu2016}. A recipient of a (scam) email or SMS text, on the other hand, even if often urged by a pretext to respond ``on the spot'' (e.g., account termination, compromised credentials), could take advantage of the \textit{non-real} time nature and look for ``cues'' of deception in the email/text or revisit the request at a later time before deciding what action to take. The transactional aspect of email and SMS communication thus allows for providers to apply extensive filtering in case of emails (e.g., ``spam'' or ``junk'' inboxes), reporting and user labeling (e.g., a ``Report Junk'' links at the bottom of any message from any unknown sender)~\cite{ApplePhishing}, and elaborate warnings about the phishing, spam, or scam nature of the requests (e.g., banners reading ``Why this message is dangerous'' or ``This message is reported junk'')~\cite{GooglePhishing, OutlookPhishing}.  

The \textit{real} time response limits, in the case of telephone scams, how much a protection could introduce latency in the natural call flow in order to authenticate a caller and investigate the nature of the call for the purpose of blocking and/or labeling. There is no advantage of ``spam'' or ``junk'' call inboxes, any reporting could only be done after the call is answered, and the only warning that a user receives is a CallerID labeled \texttt{Scam Likely}~\cite{McEachern2018}. Telephone scam call recipients enjoy no advantage of images, logos, or any multimedia that usually helps email or SMS scam recipients cue a deception. Nor can they rely on skimming the text of the request to get a better idea of what is asked from them (even if they let the phone go to voice mail and use a transcription service, the transcribed message is usually of low quality that is practically useless). And even if recipients might not trust a call (e.g., it comes from an unknown number or is labeled as \texttt{Scam Likely}), they might feel they have to answer it anyhow because it might be related to work, a personal engagement, or an expected callback.

\subsection{Telephone Scam Prevention}
Restricted as such, the recipients of telephone scams have mostly themselves to rely on for avoiding telephone scams as scammers rarely reuse phone numbers~\cite{Prasad2023}. An independent empirical evaluation of the telephone scam problem run by Tu et al. revealed that scammers could effectively retrieve users' Social Security numbers by employing a government impersonation pretext and using a ``neighbor spoofing'' form of the caller ID (formatted such as it matches the first three or six digits of the recipient's phone in order to entice them to answer the phone call believing it is a neighbor or local organization, e.g., a school)~\cite{Tu2019}. In such a situation, Edwards et al. tested a STIR/SHAKEN verified indicator -- a green circle containing the letter ``V'' and a text reading ``Verified Number'' as a label underneath the caller ID - to help users overcome the spoofed caller ID and local area code trick~\cite{Edwards2020}. While this indicator helped users decide to trust the call, it did not help them determine the context of the call (i.e., the scamming intent of the caller), leaving the opportunity for scams to hide behind verified telemarketers or any other types of legal phone spams. 

To determine the call context and spare the recipient from being bothered by the unwanted call in the first place, Pandit et. al proposed a virtual assistant that answers the call and interacts with the caller to determine the nature of the call~\cite{Pandit2023}. A user evaluation suggested that users welcome additional information about the call context and an intermediary that emulates a ``spam/junk'' filtering functionality. As the implementation of such an intermediary might be costly and introduce prolonged latency in the normal call flow for users, Du et al. proposed an option where users could set up call-blocking policies using attribute-based policies (rather than the caller ID) that offer the flexibility of defining unwanted call based on predefined call contexts~\cite{Du2023}. The solution relies on decoupled authentication and call setup processes and it was only evaluated from a latency perspective, though no user evaluation was presented about its usability and potential adoption.

Warning users about a potential scam telephone call while blending call-blocking functionality and implicit notification of the call's context is a feature promised by most of the third-party ``anti-robocall'' apps. To see whether this promise is kept, Sherman et al. evaluated the impact of interface design elements of the ten most popular ``anti-robocall'' apps on the user's ``on the spot'' decision-making ~\cite{sherman2020you}. Users involved in the evaluation positively selected the use of signage (e.g., general prohibition sign), background contrast, Caller ID verification, and the ability to monitor call logs or see app-provided statistics to determine the call context (e.g., ``this phone number was blocked by X number of other users''). Building upon these findings, Munyaka et al. tested a variant of the call context determination element where the accuracy of various telephone scam call warnings was communicated to the users~\cite{Munyaka2024}. The results of this test indicated that knowing how accurate a telephone scam warning affects the perceptions of the call context and nudges the recipient to err on the side of caution.  

% As an act of retribution against the telephone scammers Sahin et al. analyzed the role of a chatbot or a ``robocallee'' that answered robocalls on the behalf of recipients and instead of notifying them, mostly turned an automated set of conversational prompts against the scammers by the way of wasting their time and resources~\cite{Sahin2017}. 

% “verifed” call only means that the call is coming
% from the device it says it is. It says nothing about the intent
% of the caller. All the unwanted sales calls coming from valid
% numbers could also be marked as verifed.
% There are two approaches to displaying a STIR/SHAKEN
% indicator. The frst leverages the caller ID display only, so that
% a label is presented with the number. Alternatively, a“Verstat”
% indicator could be sent in the call header and recognized by the
% phone hardware so that calls can utilize graphics, iconography,
% and messaging in a robust manner. Because this must be done

\subsection{Telephone Scam Warnings Accessibility}
A large-scale analysis of scam calls using a honeypot over almost a year revealed that telephone scams target vulnerable and at-risk populations, in particular older adults or recent immigrants~\cite{Prasad2020}. A further evaluation showed that the at-risk populations are usually targeted with phone calls that use Social Security disability benefits as a pretext to engage with the call recipients~\cite{Prasad2023}. Often, at-risk populations such as older adults, immigrants, and people with disabilities don't enjoy the same degree of protection when it comes to warnings about unwanted emails because the designs usually involve heavy graphical components, technical wording, or are not accessible at all~\cite{Bellini2024, Sharevski2024-usenix}. At the same time, these at-risk populations heavily rely on the telephone as a communication service partly because of the convenience and familiarity with it, partly because telephone calls are far more flexible and accessible than email or SMS texts (e.g., for people with visual disabilities). 

Recognizing that telephone scams place at-risk populations at a protection disadvantage, Sherman et al. also did an accessibility evaluation of the 56 ``anti-robocall'' apps with a set of legally blind users~\cite{sherman2021characterizing}. None of the apps were accessible through any assistive technology, offering no minimum color contrast for low vision users, no tags and labels on buttons for managing a call for navigation, nor any automatic audible alerts. Sherman et al. also involved legally blind users in the evaluation of aural warnings that spoke back variants of the verification label applied next or instead of the Caller ID~\cite{Sherman2020a11y}. Their results suggest that legally blind users favor plain language when communicating the type of call in order to be able to avoid confusion when deciding whether to take the call ``on the spot'' or not. 

Munyaka et al. also involved legally blind users when testing the effect of telephone scam call warnings' accuracy on the ability to infer the context of the call~\cite{Munyaka2024}. Their findings indicate that visually impaired people heavily rely on Caller ID and, the verification label (e.g., \texttt{Scam Likely}) to determine how to proceed with a call and information about the accuracy adds a degree of confusion that could heighten their risk of being scammed (e.g., the blind users were confused why a Caller ID they recognized was also determined by the carrier to likely be a scam). The confusion was also a reason for Voight et al. to go beyond just the warning within the incoming call and propose a safe call-answering solution for older adults that involved an NFC card~\cite{Voigt2023}. The idea is to build up a trusted source of callers to enable an accompanying card to give a warning to incoming calls from unfamiliar callers.

\subsection{Contextual Telephone Scam Warnings}
The protection measures against telephone scams, so far, hardly enable users to infer the context of the call, that is, to enable them to go beyond the Caller ID and the verification label in order to determine how to proceed. Restricted by the brief amount of time (e.g., several seconds, during the call ringing period), the design space for cuing the exact deception context ``on the spot'' is thus limited. To overcome this limitation, alternative scam protection is to alert users about the context \textit{during} the call, i.e., after they answer it (or let it go to the voice mail, nonetheless, where the alert will also be recorded/transcribed). But such a solution requires \textit{listening} or \textit{sampling} calls for conversation patterns commonly associated with scams and, using a generative AI engine, dynamically \textit{insert} the warning. 

Phone device manufacturers and app developers could feasibly implement this as an entirely client-site solution, but we haven't seen it yet for obvious risks of centralized control and privacy invasion~\cite{Chan2024} (a more extensive discussion about the privacy implications of such a solution, including participants' insights on this issue gathered as part of this study, is offered in section~\ref{sec:privacy-discussion}). Google recently hinted at a step towards such a solution, previewing a built-in Android OS feature that uses Gemini Nano (Google's generative AI engine) to dynamically offer a contextual scam cue ``on the spot,'' even if the picked up~\cite{Heather-Google-Scam}. An example contextual warning that Android plans to insert on the phone screens immediately after a call is answered is shown in Figure \ref{fig:visualwarning}. 

\begin{figure}[h]
\centering
        \includegraphics[width=0.8\linewidth]{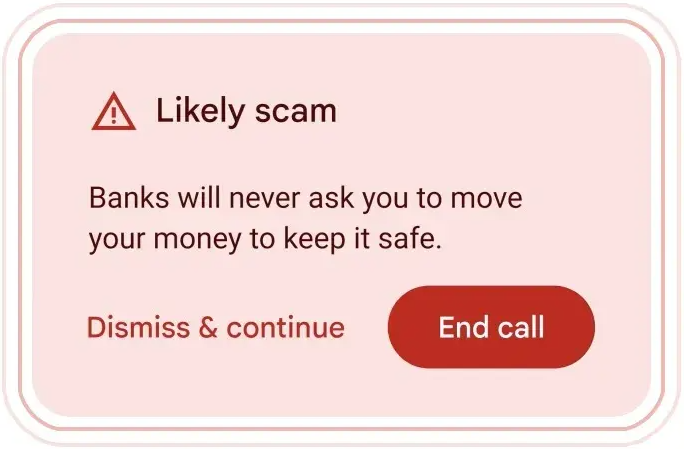}
 \caption{Contextual Telephone Scam Warning} 
\label{fig:visualwarning}   
\end{figure}

Following Wogalter et al.~\cite{Wogalter2002} warning design recommendations, this warning retains the cuing element that signals to the receiver the nature of the call (Likely scam) though in this case it's not directly associated with the CallerID like in the previous cases~\cite{Sherman2020a11y, Edwards2020}. This helps a receiver to identify the hazard (i.e., an unwanted call) ``on the spot.'' Once this happens, the warning shows the context where the receiver is notified about the type and nature of the scam, for example, ``\textit{Banks will never ask you to move your money to keep safe.}'' At this point, users are not directly offered an explanation of consequences if exposed to the hazard (i.e., loss of money), but are given just the context so they can infer the negative outcomes if they follow up with the call request (one could conjecture that this is a deliberate design choice that guards for potential false positives and actual bank calls, for example). The contextual cue about the nature of the scam thus comes close to the idea of just-in-time and just-in-place warnings implemented in the case of (scam) emails and SMS texts~\cite{Volkamer2017, Petelka2019}. Lastly, receivers are given directives for avoiding the hazard and offered the option to either ``Dismiss \& continue'' with the call (an implicit button that requires clicking on the text) or click an explicit button that will ``end call''. 

Obviously, the warning entirely relies on visual friction for a user to heed the context and avoid being scammed. A cursory accessibility evaluation similar to the one performed by Sharevski and Zeidieh for (scam) email warnings~\cite{Sharevski2024-usenix} would inevitably raise concerns about the presence of navigation elements so a screen reader could separately convey the banner, the context, and the button/action options. Following the concerns outlined by Sherman at al.~\cite{Sherman2020a11y} and Munyaka at al.~\cite{Munyaka2024}, a visually impaired receiver would perhaps find it difficult to follow up the context and intuitively make sense of its language as well as the wording of the buttons. And, specifically for legally blind people, there is the dimension of relevance to the accessibility. Even if Google makes the warning technically accessible, there is no guarantee that the warning will be relevant to a visually impaired person. For example, if the contextual warning is about auto warranty expiration, the context might read ``\textit{Car insurers will never ask you to renew over the phone},'' but driving a car is not something they usually do.

% \begin{table*}[!ht]
% % increase table row spacing, adjust to taste
% \renewcommand{\arraystretch}{1.1}
% % \footnotesize
% % \renewcommand{\tabcolsep}{2mm}
% \caption{Password Manager Usability Evaluation}
% \label{tab:background}
% \centering
% \footnotesize
% \aboverulesep=0ex % Solution part 1 of 3
%    \belowrulesep=0ex % Solution part 1 of 3
% \begin{tabularx}{\linewidth}{|cXXXXXX|}
% \toprule
% \textbf{Study} & \textbf{\# Users} & \textbf{PM Types} & \textbf{Storing} & \textbf{Generating} & \textbf{Populating} & \textbf{Security}   \\
% \midrule
% \cite{}  & & & & & & \\
% \cite{}  & & & & & & \\
% \cite{}  & & & & & & \\
% \cite{}  & & & & & & \\
% \cite{}  & & & & & & \\

% % \cite{}  & & & & & & \\

% \hline
% \end{tabularx}
% \end{table*}
% ################################
% \clearpage
\section{Study} \label{sec:study}
As the regulatory push for curbing spam would necessitate some form of contextual telephone warnings to be conveyed to users~\cite{fsra-2022}, we wanted to explore how these warnings ``work'' in an accessible variant i.e., as \textit{aural} warnings that pertain to both visually impaired and sighted users. We refrained from just an accessibility evaluation of Google's warning shown in Figure \ref{fig:visualwarning} because we were more interested in how users -- both legally blind and sighted -- would react to relevant, aural contextual scam warnings in \textit{naturalistic}, rather than in laboratory settings. We obtained an expedited Institutional Review Board (IRB) approval for a mild deception study that involved 36 legally blind\footnote{Legally blind individuals in the US include those with acuity of 20/200 or field-of-view of 20 degrees or less in the better eye with correction; low vision with acuity up to 20/70 and field-of-view larger than 20 degrees in the better eye with correction~\cite{statutory-blindness}.} and 36 sighted adults from the US that have received scam calls in the past (the protocol-related information including the research safeguards is elaborated in section~\ref{sec:recrutiment}). To obtain first-hand experiences with contextual scam warnings we first performed a pilot study to finalize the study stimuli and adequately situate the main study into the naturalistic setting experiences by both legally blind and sighted people targeted by telephone scam calls.

\subsection{Pilot Study}
After we obtained consent, we created an aural translation of Google's contextual warning using both a feminine and masculine stock synthetic voice from the ElevenLabs library of synthetic voices~\cite{elevenlabs}. As we couldn't exactly replicate the ``Dismiss \& continue'' and ``End Call'' buttons, we experimented with an (1) ``action'' variant using the standard phone keypad buttons (press one to continue, press zero to end the call); and a (2) ``narrative'' variant where we simply asked the users to stay on the line or hang up using the standard ``End Call'' button. We conducted a pilot study with two legally blind and two sighted participants to test the aural and contextual variants (we obtained the initial IRB approval with the exact aural translation of the warning shown in Figure~\ref{fig:visualwarning} and we amended the protocol later on with the final stimuli before we commenced with the main study). We also used the pilot study to present participants with the most prevalent telephone scams per the FTC~\cite{ftc-scams} and decide on the two most relevant real-world scenarios as we wanted not just to study the contextual warning offered by Google~\cite{Heather-Google-Scam} associated with the FTC's example for \textit{interest rate} scams, but at least an additional one to ensure the robustness of our findings. 

During the pilot study, all the participants agreed to proceed with the second variant where no action is asked from the receiver as that part of the warning might be mistakenly confused as part of the scam itself. Given that many legally blind people in the US receive Social Security benefits, our pilot participants suggest that we use the FTC's example for the \textit{Social Security} scam as all of them have received similar scam calls in the past. As the aural translation of the contextual warnings might run long, our pilot participants suggested we test a short warning variant that doesn't carry the context per se, but nonetheless urges the receivers to hang up on the call or proceed with caution. The resultant choice of scams, including the contextual warnings and the short warnings, are given in Tables \ref{tab:interest-rate-scam} and \ref{tab:social-security-scam}. 

Through the pilot study deliberation, our legally blind participants suggested a slight rewording of the contextual warning to avoid the ambiguities present in the one proposed by Google~\cite{Heather-Google-Scam} such as stating that ``\textit{This call is likely a scam}'' (not just ``\textit{Likely scam}'') and pointing to the scarcity principle of social engineering influence~\cite{cialdini2001science} by ``\textit{applying time pressure to make account changes}'' instead of the just one resultant action -- ``\textit{move money}.'' They suggested to alter the tone between the warning and the scam voice. Accordingly, we used a feminine warning voice for the \textit{interest rate reduction} scam (originally a masculine voice) and vice versa, a masculine voice for the \textit{Social Security} scam.

\begin{table}[!h]
\renewcommand{\arraystretch}{1.2}
% \footnotesize
% \renewcommand{\tabcolsep}{2mm}
\centering
\small
\aboverulesep=0ex % Solution part 1 of 3
   \belowrulesep=0ex % Solution part 1 of 3
\caption{Interest Rate Scam Scenario}
\label{tab:interest-rate-scam}
\begin{tabularx}{\linewidth}{|Y|}
\toprule
\textbf{Short Aural Warning} \\\hline
This call is likely a scam. Hang up immediately, otherwise, proceed with caution. \\\hline

\textbf{Contextual Aural Warning} \\\hline 
This call is likely a scam. Banks will never time pressure you to make any change to your account over the phone. hang up immediately, otherwise, proceed with caution. \\\hline

\textbf{Scam~\cite{ftc-scams}} \\\hline

\textit{Thanks to your good payment history and good credit score, you have been qualified finally for interest rate reduction between 0 to 5\%. Several attempts were made to reach you. This is your final courtesy call before we are unable to lower your interest rates, so press one now.}
\\\hline

\end{tabularx}
\end{table}

\begin{table}[!h]
\renewcommand{\arraystretch}{1.2}
% \footnotesize
% \renewcommand{\tabcolsep}{2mm}
\centering
\small
\aboverulesep=0ex % Solution part 1 of 3
   \belowrulesep=0ex % Solution part 1 of 3
\caption{Social Security Scam Scenario}
\label{tab:social-security-scam}
\begin{tabularx}{\linewidth}{|Y|}
\toprule
\textbf{Short Aural Warning} \\\hline
This call is likely a scam. Hang up immediately, otherwise, proceed with caution. \\\hline

\textbf{Contextual Aural Warning} \\\hline 
This call is likely a scam. Government agencies will never ask about your Social Security number or discuss legal matters over the phone. Hang up immediately, otherwise, proceed with caution. \\\hline

\textbf{Scam~\cite{ftc-scams}} \\\hline

\textit{We are calling you from the Department of Social Security Administration. The reason that you received this phone call from our department is to inform you that there is a legal enforcement action filed on your Social Security number for fraudulent activity, so when you get this message, kindly press one to connect with the next available officer.} 
\\\hline

\end{tabularx}
\end{table}

\subsection{Main Study Setup}
Once we finalized the study stimuli shown in Tables \ref{tab:interest-rate-scam} and \ref{tab:social-security-scam}, we structured a 2 x 3 x 2 study in naturalistic settings where two groups of participants (legally blind/sighted) were cold-called with one of the three possible study conditions (baseline, short warning, and contextual warning) in each of the real-world telephone scam cases (interest rate/Social Security). Situating the study in naturalistic settings, we had to ensure our participants were not exposed to greater than minimal risk while we cold-called them so we could study the natural response to real-world telephone scam calls and obtain consent afterward. To do so, we first recruited two groups of participants, legally blind and sighted. For the legally blind group, we followed Gerber's advice when doing usability and accessibility studies with visually disabled people~\cite{Gerber2002} and recruited individuals 18 years of age or older who have encountered unwanted calls in the past, use assistive technology, and are English-speaking and literate. We did a snowball sampling where we partially sampled personal acquaintances who are legally blind and partially a pool of legally blind participants that was recommended by one of our acquaintances. For the sighted group, we used the same inclusion criteria (sans the use of assistive technologies criterion) only we did random sampling through Prolific.

Technically we had six study conditions, and we reasonably balanced for a random assignment of six legally blind or sighted participants per each condition (a total of 36 legally blind and 36 sighted participants). While our goal was to balance the conditions, we nonetheless made a particular effort to representatively sample each of the groups per the gender, race/ethnicity, and age dimensions (see section \ref{sec:data-collection-analysis}). We used a formal email approved by our IRB (see Appendix \ref{sec:recruit}) to approach each of the potential participants. The email invited our participants to join a 45-minute, audio-only, recorded Zoom interview session to discuss general experiences with unsolicited online communication. We used this deception pretext to remain general and not tip the participants off about the nature of the study. The only stipulation we had was that they use Calendly to sign up for the study and provide their phone number together with their email so we could send them reminders for their study appointment. 

We then used their phone number to cold-call them 24 hours before their participation time was scheduled (we initially built a buffer in the Calendly to enable us to do so on a rolling basis for all participants). We decided to do the call a day earlier to ensure participants don't suspect anything is amiss or the call might be related to the study. Following the approach implemented by Tu et al.~\cite{Tu2019}, we used CallFire to create six Interactive Voice Response (IVR) campaigns that were initiated from an innocuous phone number we purchased for the purpose of the study~\cite{CallFire}. We set up the IVR logic to record if a participant presses the number one indicated in the scams in order for us to observe the natural action and the effect of the warnings (or the absence of them). Tu et al.~\cite{Tu2019} asked for the unwitting participants to enter four digits corresponding to the last four digits of their Social Security number, but we refrained from doing so because (i) our goal was to see if the participants would do as the original real-world scam requests (press one); and (ii) we deemed the collection of anything related to Social Security numbers in the context of our legally blind participants who do receive Social Security benefits too risky. If a participant pressed one (or any other number, for that matter), they were automatically transferred back to a phone number controlled by the researchers where we had the ability to speak to them. At this point, the participants were debriefed (the entire debriefing about this deception is given in Appendix \ref{sec:debrief}), consent was obtained, and we asked them to join during their scheduled time a day later. 

If they didn't do anything or let the call go to voice mail, we did the debriefing at the start of the scheduled interview and obtained consent for retaining their data from the study. We were aware that the classification of untrustworthy phone calls was predicated on the individual's telephone service provider, and we expected that we might encounter cases where the cold-call might not end up going through to the participant. We were also aware that we couldn't control for any STIR/SHAKEN CallerID warnings (e.g., \texttt{Scam Likely}) presented by the participants' telephone service provider (or for that matter, any app-based scam blocker, for example, Hiya) but we accepted these cases as part of the naturalistic settings.

\subsection{Data Collection and Analysis} \label{sec:data-collection-analysis}
To collect data, we arranged audio-only recorded Zoom interviews with interested respondents on a rolling basis that lasted on average 45 minutes, and we compensated each participant with a \$20 Amazon eGift Card (\$1,520 in total, accounting for the compensation of the pilot participants too). Initially, the interview transcripts from our Zoom sessions were not anonymized, but we removed any names and references to individual participants and deleted the audio recordings altogether. The transcripts, assigned only with a participant number in the order of participation, were stored on a secure server that only the researchers had access to. Each interview was done with open-ended questions, listed in the interview script (see Appendix \ref{sec:script}). For our sighted participants, we offered the option for them to share their perspectives about the original visual warning from Figure~\ref{fig:visualwarning}. 

We concluded our recruitment with a sample of 72 (36 legally blind and 36 sighted) participants as we reached thematic/data saturation (i.e., we collected data up to the point where there were fewer surprises in the responses to the research questions and no more emergent patterns). The demographics, call setup, and the sample's visual profile, per the suggestion in \cite{Gerber2002}, are all given in Tables \ref{tab:demographics-blv} and \ref{tab:demographics-sighted}.

\begin{table}[!h]
\renewcommand{\arraystretch}{1.1}
% \footnotesize
% \renewcommand{\tabcolsep}{2mm}
\centering
\small
\aboverulesep=0ex % Solution part 1 of 3
   \belowrulesep=0ex % Solution part 1 of 3
\caption{BLV Participant Demographic Distribution}
\label{tab:demographics-blv}
\begin{tabularx}{\linewidth}{|Y|}
\hline
% \toprule
 \textbf{Gender} \\\hline
% \midrule
\vspace{0.2em}
    \hfill \makecell{Female \\ 22} 
    \hfill \makecell{Male \\ 13} 
     \hfill \makecell{\textbf{Non-Binary} \\ 1} \hfill\null
\vspace{0.2em}
\\\hline

\textbf{Racial/Ethnic Self Identification} \\\hline
\vspace{0.2em}
    \hfill \makecell{White \\ 22} 
    \hfill \makecell{Latinx \\ 5} 
    \hfill \makecell{Asian \\ 4} 
    \hfill \makecell{Black \\ 2}
    \hfill \makecell{Other \\ 3} \hfill\null
\vspace{0.2em}
\\\hline
\textbf{Age} \\\hline
\vspace{0.2em}
\hfill \makecell{[18-29]\\ 9} 
    \hfill \makecell{[30-39] \\ 12} 
    \hfill \makecell{[40-49]\\ 9} 
    \hfill \makecell{[50-59] \\ 5}
    \hfill \makecell{[60+] \\ 1} \hfill\null
\vspace{0.2em}
\\\hline
 \textbf{Visual Self Identification} \\\hline
\vspace{0.2em}
    \hfill \makecell{Totally Blind \\ 18}
    \hfill \makecell{Low Vision \\ 18} \hfill\null
\vspace{0.2em}
\\\hline
 \textbf{Provider} \\\hline
 \vspace{0.2em}
    \hfill \makecell{AT\&T\\ 3}
    \hfill \makecell{Spectrum\\ 2}
    \hfill \makecell{T-Mobile\\ 11}
    \hfill \makecell{Verizon\\ 10}
    \hfill \makecell{Other\\ 10} \hfill\null
    % \hfill \makecell{\textbf{Windows PC}\\ 4 (14\%)} 
\vspace{0.2em}
\\\hline

\end{tabularx}
\end{table}

\begin{table}[!h]
\renewcommand{\arraystretch}{1.1}
% \footnotesize
% \renewcommand{\tabcolsep}{2mm}
\centering
\small
\aboverulesep=0ex % Solution part 1 of 3
   \belowrulesep=0ex % Solution part 1 of 3
\caption{Sighted Participant Demographic Distribution}
\label{tab:demographics-sighted}
\begin{tabularx}{\linewidth}{|Y|}
\hline
% \toprule
 \textbf{Gender} \\\hline
% \midrule
\vspace{0.2em}
    \hfill \makecell{Female \\ 19} 
    \hfill \makecell{Male \\ 17} \hfill\null
\vspace{0.2em}
\\\hline

\textbf{Racial/Ethnic Self Identification} \\\hline
\vspace{0.2em}
    \hfill \makecell{White \\ 24} 
    \hfill \makecell{Latinx \\ 3} 
    \hfill \makecell{Asian \\ 2} 
    \hfill \makecell{Black \\ 4}
    \hfill \makecell{Other \\ 3} \hfill\null
\vspace{0.2em}
\\\hline
\textbf{Age} \\\hline
\vspace{0.2em}
\hfill \makecell{[18-29]\\ 5} 
    \hfill \makecell{[30-39] \\ 7} 
    \hfill \makecell{[40-49]\\ 10} 
    \hfill \makecell{[50-59] \\ 7} \hfill\null
    \hfill \makecell{[60+] \\ 7} \hfill\null
\vspace{0.2em}
\\\hline
 \textbf{Provider} \\\hline
 \vspace{0.2em}
    \hfill \makecell{AT\&T\\ 6}
    \hfill \makecell{Spectrum\\ 2}
    \hfill \makecell{T-Mobile\\ 7}
    \hfill \makecell{Verizon\\ 15}
    \hfill \makecell{Other\\ 6} \hfill\null
    % \hfill \makecell{\textbf{Windows PC}\\ 4 (14\%)} 
\vspace{0.2em}
\\\hline
\end{tabularx}
\end{table}

Since we had to work with a degree of arbitrary selection of calls due to the natural settings, we asked our participants to provide lengthy responses to our questions and asked for further clarifications. With the collected data, we performed an inductive coding approach to identify frequent, dominant, or significant aspects of their answers. As suggested in \cite{clarke2015thematic}, we first familiarized ourselves with the data and then we completed a round of open coding for arbitrarily selected two interviews to capture the participants' actions and reflections. Then we discussed the individual coding schemes and converged on an agreed codebook (see Appendix~\ref{sec:codebook}). Two researchers used this codebook to individually code all interview transcripts. Whenever necessary, they added new codes for concepts not yet covered by the codebook (six in total). Afterward, we compared the results and resolved disagreements in the code assignments. We specifically did not include inter-rater-reliability calculations in our process, as our goal was to freely explore diverse topics associated with the natural behavior around telephone scam calls (coodebook thematic analysis). Overall, our process follows the best practices of usable security and privacy research outlined by Ortloff et. al~\cite{ortloff2023}. Finally, we discussed and interpreted the identified themes respective to our research questions and wrote down the results. For this, we selected example quotations to represent each of the findings \cite{fereday2006demonstrating}.

% The codebook (see Appendix \ref{sec:codebook}), captured three main aspects: (i) \textit{initial response} i.e., codes pertaining to the initial response to the study phone calls, some of which included contextual scam warnings; (ii) \textit{reflection} i.e., codes pertaining to the response upon reflection about the study phone calls, some of which included contextual scam warnings; and (iii) \textit{contextual telephone scam warnings} i.e., codes pertaining to the reception of the scam warnings used in the study.

In reporting the results, we utilized verbatim quotation of the participants' answers as much as possible, emphasized in ``\textit{italics}'' and with a reference to the participant as either \textbf{PV\#\textsubscript{MW}} or [\textbf{PV\#\textsubscript{MW}}], where \textbf{P} denotes \textbf{participant}, \textbf{V} denotes the \textbf{vision ability level} of the participant (\textbf{B} - legally blind, \textbf{S} - sighted), and \textbf{\#} denotes the order of participation. For each of the conditions \textbf{M} denotes the \textbf{scam} message they received (\textbf{F} - Social Security fraud, \textbf{I} - interest rate reduction), and \textbf{W} denotes the \textbf{warning} they heard (\textbf{B} - baseline, \textbf{S} - short warning, \textbf{C} - contextual warning. For example, \textbf{PB2\textsubscript{FS}} denotes the second participant, who is legally blind and was cold-called using the Social Security fraud scam preceded by the short aural warning.

\subsection{Trust and Ethical Considerations} \label{sec:recrutiment}
As this was a study in naturalistic settings and concerning the participants' \textit{own} phone numbers, it was important for us to establish trust, provide assurances, consider the accessibility needs, and offer support throughout the participation. Following the suggestions for involving at-risk populations in cybersecurity research~\cite{Sharevski-IEEEMag} we obtained verbal consent, following the debriefing about the nature of the call. We communicated that the goal of our study was to capture the ``richness'' of their natural experiences with telephone scam calls rather than the commonality of these experiences in laboratory settings. This is the reason, hence, why we called their \textit{own} phone number and, if applicable, their \textit{own} access technology. We offered emotional and accessibility support on an ongoing basis during the interview because talking about telephone scam calls might be distressing. We thus allowed the participants to take breaks if needed (recording stopped) and then continue if they wished or, reschedule the interview at a later time of their better convenience.

% Before and during the interviews, at all times, we gauged whether or not any potential participant was too vulnerable to take part in the first place. We did not have such an instance, but our plan was to offer them an alternative time and date for the interview, in addition to the other means of support. 

% both before we started the audio recording and afterward (to have evidence in our transcripts, but also to avoid creating a recording in case a potential participant does not consent to keep their record after the debriefing, in which we would have thank them, delete any entry related to them, and closed the Zoom session). All of the participants in both groups agreed to keep their records. 

Only after we received the participants' explicit permission that they were okay with proceeding with the study and discussing their experience with the cold call we placed a day before, we commenced the audio-only recorded Zoom session and proceeded with the interviews. We allowed them to verbalize the process, give comments, complaints, suggestions, and verbalize any other experience that was not necessarily with unwanted calls but other technologies (e.g., scam, spam, or phishing emails/SMS texts) in order to allow for them to fully express the natural behavior ``surrounding'' their daily interactions with unsolicited communications. 

The consent process made it clear that the participation is anonymous, i.e., no names, addresses, phone numbers, or any personally identifiable information is collected. Participants were informed that they were free to stop and abandon any question at any point in time, remove any answers or withdraw from the study either immediately after finding out about the mild deception (i.e., the cold call a day before) or up to 30 days after they participated. To attempt to locate and remove their data we offered the participants to recall a unique way to identify their answers (no one requested removal or declined the participation). We also pointed out to our participants that they could act on the phone call from the study as they ultimately wished (e.g., delete a voice mail, add it to a scam call-blocking app, etc.). We verbosely debriefed them about the mild deception we used and that we were the ones who initiated the scam-sounding cold call to their phone number. We offered aditional telephone scam protections resources if they wished to further raise or check their awareness \cite{ftc-scams}. We reviewed the main points we recorded during the interview and clarified any misunderstandings we might have. We also sent a draft of our paper to our participants for feedback.

% Though anonymity was strictly safeguarded, we offered the option for the participants to share a contact to break confidentiality if they or we believed there to be a risk of harm, a need for immediate assistance, or an emergency (no such instance occurred).

We employed lengthy explanations to ensure participants that we were not involved with the way scam calls are handled or otherwise processed by telephone service providers, phone device manufacturers, or app developers. We also notified participants that we are not involved with the design and implementation of the contextual warning that we based our study on. We were also careful not to appear in favor nor support of particular types of warnings in order to maintain full researcher impartiality. We communicated that our ultimate goal is to meaningfully \textit{include} legally blind individuals in cases where contextual warnings are otherwise visually available for sighted people. We pointed out that, this goal, however, doesn't prevent from misusing our findings or misinterpreting them in making compromises for accessibility or removing such support altogether.

\section{Results}
% KEY: 
% LABEL: P - PARTICIPANT
% NUMBER:  1 - ORDER OF PARTICIPATION 
% VISUAL DIAGNOSIS: X - BLIND ; Y - SIGHTED 
% TYPE OF SCAM: A - SOCIAL SECURITY ; R - INTEREST RATE REDUCTION 
% TYPE OF WARNING: B - BASELINE; S- SHORT; L - LONG

\subsection{RQ1a: Interest Rate Scenario}
% \item \textbf{RQ1a:} How would \textit{legally blind} people initially respond to an  \textit{interest rate reduction} telephone scam, compared to their \textit{sighted} counterparts, in three conditions: (1) \textit{without} a warning (baseline); (2) with a \textit{short aural} warning; and (3) with a \textit{contextual aural} warning?

\subsubsection{Initial Response}
% Q1
The initial response to the day-before cold-call for the \textit{interest rate reduction} scenario (Table~\ref{tab:interest-rate-scam}) for each of the participant groups is given in Table~\ref{tab:initial-response-irr}. We extracted call data records from CallFire to determine the action and call duration. Participants acted naturally as they do with all other types of telephone scams as we called them from an innocuous looking number. None of the participants in either group ``pressed one'' as requested in the original text of the \textit{interest rate reduction} scam.  Overall, the participants in the \textit{legally blind} group mostly let the call go to ``voice mail'' (eight), hung up after listening to the warning (four), or simply hung up (three)/declined (three). The \textit{sighted} participants either let the call go to ``voice mail'' (twelve) or hung up after listening to the warning (six). 

Per condition, all of the \textit{sighted} participants let the baseline scam call go to voice mail while only two of the \textit{legally blind} did so. Three of them picked up, listened to the scam message, and hung up and one of them declined the call. Two participants in each group listened to the \textit{short warning} and hung up the call, while the rest of them let the call go to voice mail (one \textit{legally blind} participant declined the call). Two \textit{legally blind} participants listened to the \textit{contextual warning} scenario (one declined it), but four of their \textit{sighted} counterparts did so (the rest let the call go to voice mail).

\begin{table}[!h]
\renewcommand{\arraystretch}{1.1}
% \footnotesize
% \renewcommand{\tabcolsep}{2mm}
\centering
\small
\aboverulesep=0ex % Solution part 1 of 3
   \belowrulesep=0ex % Solution part 1 of 3

\caption{\textbf{Initial Response} -- \textit{Interest Rate} Scenario}
\label{tab:initial-response-irr}
\begin{tabularx}{\linewidth}{|YYY|}
\toprule

\textbf{Baseline} & \textbf{Short} & \textbf{Contextual} \\\hline

\multicolumn{3}{|c|}{\textbf{Legally Blind Participants}} \\\hline
Voice Mail & Hang Up$^*$ & Voice Mail \\
Hang Up & Voice Mail & Hang Up$^*$ \\   
Voice Mail & Voice Mail & Voice Mail \\
Hang Up & Declined & Voice Mail \\   
Declined & Hang Up$^*$ & Declined \\
Hang Up & Voice Mail & Hang Up$^*$ \\\hline   

\multicolumn{3}{|c|}{\textbf{Sighted Participants}} \\\hline
Voice Mail & Voice Mail & Voice Mail \\
Voice Mail & Hang Up$^*$ & Hang Up$^*$ \\
Voice Mail & Voice Mail & Hang Up$^*$ \\
Voice Mail & Hang Up$^*$ & Hang Up$^*$ \\
Voice Mail & Voice Mail & Hang Up$^*$ \\
Voice Mail & Voice Mail & Voice Mail \\\hline   

\multicolumn{3}{|l|}{$^*$ indicates hanging up \textit{after} listening to the warning} \\\hline

\end{tabularx}
\end{table}

\subsubsection{Scam Event Recall}
% Q2
After the participants joined the audio-only Zoom interview, we debriefed them about the nature of the cold call that we placed a day before, obtained consent to keep their data, and asked them to recall how they acted on it. Table~\ref{tab:action-recall-irr} breaks down the cues that each participant used to decide what action to take ``on the spot.'' Per condition, the \textit{legally blind} participants that answered but hung up decided to do so either because they didn't know the number or they saw a STIR/SHAKEN indicator \texttt{Scam Likely}. This indicator was sufficient for one of them to directly decline the call. One of the \textit{legally blind} participants that let the call go to voice mail said they don't check their inbox and one of them said they listened to it but the unknown number was a sufficient cue not to do anything. As all of the \textit{sighted} participants let the call go to voice mail in the baseline condition, they decided not to act on it either because they didn't know the number or heard the keyword ``\textit{press one}'' which was a cue that this is a scam. 

\begin{table}[!h]
\renewcommand{\arraystretch}{1.1}
% \footnotesize
% \renewcommand{\tabcolsep}{2mm}
\centering
\small
\aboverulesep=0ex % Solution part 1 of 3
   \belowrulesep=0ex % Solution part 1 of 3

\caption{\textbf{Action Recall} -- \textit{Interest Rate} Scenario}
\label{tab:action-recall-irr}
\begin{tabularx}{\linewidth}{|YYY|}
\toprule

\textbf{Baseline} & \textbf{Short} & \textbf{Contextual} \\\hline

\multicolumn{3}{|c|}{\textbf{Legally Blind Participants}} \\\hline

Unknown number & Warning, synth & Warning$^{\dagger}$, unkn \\
Unknown number & Warning$^{\dagger}$ & Warning, synth  \\   
No Recall$^*$ & Warning$^{\dagger}$, unkn & Warning$^{\dagger}$ \\
\texttt{Scam Likely} & Unknown number & \texttt{Scam Likely} \\   
\texttt{Scam Likely}  & Unknown number & Unknown number \\
Unknown number & Warning$^{\dagger}$ & Warning \\\hline   

% Voice Mail & Hang Up$^*$ & Voice Mail \\
% Hang Up & Voice Mail & Hang Up$^*$ \\   
% Voice Mail & Voice Mail & Voice Mail \\
% Hang Up & Declined & Voice Mail \\   
% Declined & Hang Up$^*$ & Declined \\
% Hang Up & Voice Mail & Hang Up$^*$ \\\hline   

\multicolumn{3}{|c|}{\textbf{Sighted Participants}} \\\hline

Unknown Number & Unknown Number & \texttt{Scam Likely}  \\
Unknown Number & Warning & Warning \\
Press One$^{\dagger}$ & Unknown Number & Warning \\
Press One$^{\dagger}$ & Warning & Warning \\
Unknown Number & \texttt{Scam Likely} & Warning \\
Unknown Number & Warning$^{\dagger}$, unkn & Warning$^{\dagger}$ \\\hline   

% Voice Mail & Voice Mail & Voice Mail \\
% Voice Mail & Hang Up$^*$ & Hang Up$^*$ \\
% Voice Mail & Voice Mail & Hang Up$^*$ \\
% Voice Mail & Hang Up$^*$ & Hang Up$^*$ \\
% Voice Mail & Voice Mail & Hang Up$^*$ \\
% Voice Mail & Voice Mail & Voice Mail \\\hline   

\multicolumn{3}{|l|}{$^*$ indicates \textit{no} voice mail check or an immediate \textit{deletion}} \\
\multicolumn{3}{|l|}{$^{\dagger}$ indicates no action upon listening to the voice mail} \\

\hline

\end{tabularx}
\end{table}

In the \textit{short warning} scenario, the \textit{legally blind} participants that picked up the call, heard the warning and decided to ``\textit{hang up immediately}'' [\textbf{PB14\textsubscript{IS}}] as the warning urged them to do so. The participants in this group that let it go to voice mail, decided not to do anything because ``\textit{the message said the call was likely a scam}'' [\textbf{PB5\textsubscript{IS}}] The one participant that declined, did so because they didn't recognize the number. The \textit{sighted} participants that answered the call, equally, decided to hang up because the warning urged them to do so. Those who let the call go to voice mail indicated they didn't take any action because either they didn't recognize the number (three) or they saw an indicator \texttt{Scam Likely} on the recording. 

In the \textit{contextual warning scenario}, the \textit{legally blind} participants that picked up the call, decided to hang up both because the warning let them know the call was scam but also that they felt the actual scam message sounded ``\textit{synthetic}'' or machine-generated (a robocall). Those who let it go to voice mail, upon inspection, decided to ignore the message because the warning said it was ``\textit{likely a scam}.'' The one participant that declined the call, did so because they didn't recognize the number. All of the sighted participants who answered the call heeded the contextual warning and decided to hang up. One of those that let it go to voice mail said they didn't act on it because it had a \texttt{Scam Likely} indicator [\textbf{PS42\textsubscript{IC}}] and one of them because the warning ``\textit{explained why they shouldn't}'' [\textbf{PS45\textsubscript{IC}}].

%%%%%%%%%%%%%%%%%%%%%%%%%%%%%%%%%%%%%%%%%%%%%%%%%%%%%%%%%%%%
%%%%%%%%%%%%%%%%%%%%%%%%%%%%%%%%%%%%%%%%%%%%%%%%%%%%%%%%%%%% 

\subsection{RQ1b: Social Security Scenario}
% \item \textbf{RQ1b:} How would \textit{legally blind} people initially respond to a  \textit{social security} telephone scam call, compared to their \textit{sighted} counterparts,  in three conditions: (1) \textit{without} a warning (baseline); (2) with a \textit{short aural} warning; and (3) with a \textit{contextual aural} warning?

\subsubsection{Initial Response}
% Q1
The initial response, per the CallFire logs, to the day-before cold-call for the \textit{Social Security} scenario (Table~\ref{tab:social-security-scam}) for each of the participants groups is given in Table~\ref{tab:initial-response-ss}. Only two participants in the \textit{legally blind} group ``pressed one'' as requested in the original text of the \textit{interest rate reduction} scam, despite the presence of the long, contextual warning. Overall, the participants in this group mostly let the call go to ``voice mail'' (seven), hung up after listening to the warning (four), or simply hung up (four)/declined (one). The \textit{sighted} participants either let the call go to ``voice mail'' (fourteen), hung up after listening to the warning (three), or declined the call (one). 

Per condition, all of the \textit{sighted} participants let the baseline scam call go to voice mail bar one (declined), while only two of the \textit{legally blind} did so. The other four picked up, listened to the scam message, and hung up. Two participants in the \textit{sighted} group and three in the \textit{legally blind} group listened to the \textit{short warning} and hung up the call, while the rest of them let the call go to voice mail (one \textit{legally blind} participant declined the call). All but one \textit{sighted} participant let the call go to voice mail in the \textit{contextual warning} scenario (the one answered and hung up). Six \textit{legally blind} participants listened to the \textit{contextual warning} scenario -- one hung up, two pressed one, and the three others let the call go to voice mail.

\begin{table}[!h]
\renewcommand{\arraystretch}{1.1}
% \footnotesize
% \renewcommand{\tabcolsep}{2mm}
\centering
\small
\aboverulesep=0ex % Solution part 1 of 3
   \belowrulesep=0ex % Solution part 1 of 3

\caption{\textbf{Initial Response} -- \textit{Social Security} Scenario}
\label{tab:initial-response-ss}
\begin{tabularx}{\linewidth}{|YYY|}
\toprule

\textbf{Baseline} & \textbf{Short} & \textbf{Contextual} \\\hline

\multicolumn{3}{|c|}{\textbf{Legally Blind Participants}} \\\hline

Voice Mail & Declined & Voice Mail \\
Hang Up & Hang Up$^*$ & Voice Mail \\
Voice Mail & Voice Mail & Pressed One \\   
Hang Up & Hang Up$^*$ & Hang Up$^*$ \\
Hang Up & Hang Up$^*$ & Voice Mail \\   
Hang Up & Voice Mail & Pressed One \\\hline

\multicolumn{3}{|c|}{\textbf{Sighted Participants}} \\\hline

Voice Mail & Hang Up$^*$ & Voice Mail \\  
Voice Mail & Voice Mail & Voice Mail \\
Voice Mail & Hang Up$^*$ & Voice Mail \\
Declined & Voice Mail & Hang Up$^*$ \\
Voice Mail & Voice Mail & Voice Mail \\
Voice Mail & Voice Mail & Voice Mail \\\hline 

\multicolumn{3}{|l|}{$^*$ indicates hanging up \textit{after} listening to the warning} \\\hline

\end{tabularx}
\end{table}

\subsubsection{Scam Event Recall}
% Q2
After the debriefing in the main interview, we set to learn more about the participants' initial actions, particularly the two \textit{legally blind} participants who pressed one despite hearing the \textit{contextual warning} (both of them were debriefed when they pressed one as the call was routed back to the researcher). Per Table~\ref{tab:action-recall-ss}, \textbf{PB18\textsubscript{FC}} \textit{accidentally} pressed one because of incorrect voice-over navigation, and the  \textbf{PB9\textsubscript{FC}} did so \textit{deliberately}, reasoning: 

\begin{quote}
    ``\textit{Someone was trying to stop me from pressing, you know, button number one (the warning). But I thought, I'm going to press it, I'm going to keep the person on the phone because if they're talking to me, they're not talking to some poor little old lady down the street who's going to give them their Social Security number}''
\end{quote}

\begin{table}[!h]
\renewcommand{\arraystretch}{1.1}
% \footnotesize
% \renewcommand{\tabcolsep}{2mm}
\centering
\small
\aboverulesep=0ex % Solution part 1 of 3
   \belowrulesep=0ex % Solution part 1 of 3

\caption{\textbf{Action Recall} -- \textit{Social Security} Scenario}
\label{tab:action-recall-ss}
\begin{tabularx}{\linewidth}{|YYY|}
\toprule

\textbf{Baseline} & \textbf{Short} & \textbf{Contextual} \\\hline

\multicolumn{3}{|c|}{\textbf{Legally Blind Participants}} \\\hline

Synthetic Voice & \texttt{Scam Likely} & Synthetic Voice \\
Synthetic Voice &  \texttt{Scam Likely} & Unknown Number  \\
Unknown Number & Warning & Scam Sounding \\   
Area Code & Warning & Area Code \\
Area Code & Area Code & Synthetic Voice \\   
Unknown Number & Area Code & Warning \\\hline  

\multicolumn{3}{|c|}{\textbf{Sighted Participants}} \\\hline

Unknown Number & Warning & Warning$^{\dagger}$ \\  
\texttt{Scam Likely} & Unknown Number & Warning$^{\dagger}$ \\
Unknown Number & Warning & Unknown Number \\
No Recall$^*$ & Unknown Number & Unknown Number \\
Scam Sounding & Area Code & \texttt{Scam Likely} \\
Area Code & Scam Sounding & Unknown Number \\\hline

\multicolumn{3}{|l|}{$^*$ indicates \textit{no} voice mail check or an immediate \textit{deletion}} \\
\multicolumn{3}{|l|}{$^{\dagger}$ indicates no action upon listening to the voice mail} \\\hline

\end{tabularx}
\end{table}

The remaining \textit{legally blind} participants in the \textit{contextual warning} scenario didn't engage with the voice mail because the voice sounded synthetic or hung up after suspecting an unknown area code. The area code as cue appeared in the \textit{short warning} condition too as two of the \textit{legally blind} participants were either suspicious of it and hung up or were reluctant to act on a voice mail. The remaining participants in this group hung up heeding the warning or avoided the call as they heard the STIR/SHAKEN label \texttt{Scam Likely}. The \textit{sighted} participants that picked up the phone also heeded the warning, and those that let it go to voice mail suspected either an unknown number, a scam sounding text, or a ``scammy'' area code. 

In the baseline scenario, the \textit{legally blind} participants hang up or let the call go to voice mail because they suspected an unknown area code, unknown number, or the message sounded synthetic. The \textit{sighted} participant in the baseline scenario that declined the call had no recollection of it and the remaining one let it go to voice mail because of a suspicious area code or unknown number, acting not because, upon inspection, the message sounded synthetic or was labeled as \texttt{Scam Likely}. The area code as a cue, interestingly, appeared in the \textit{Social Security} scenario perhaps because the scam usually targets people that collect Social Security/disability benefits. We noticed that the same pretext appeared in the work done by Munyaka et al.~\cite{Munyaka2024} and our participants, as did theirs, confirmed that ``\textit{a lot of Social Security junk comes from Florida}'' so they always look suspicious for Florida area codes.

% breaks down the cues that each participant used to decide what action to take ``on the spot.'' 

\subsection{RQ2: Response and Telephone Scam Cues}
% Q3 and Q4 and Q6 and Q12
% RQ2: How do (a) legally blind and (b) sighted people usually detect and deal with telephone scam calls?
We dug deeper into the cues used by our participants to detect scam calls. The first thing people were looking at was whether the number calling them was known to them or not. 52 participants used this as an indicator when dealing with unsolicited calls (23 \textit{legally blind} and 29 \textit{sighted}). Out of these, 15 stated that they never pick up calls from unfamiliar numbers (seven \textit{legally blind}, eight \textit{sighted}), three more only when they were expecting such a call, and two who would decide based on the area code. One \textit{legally blind} and two \textit{sighted} participants explained that they would pick up calls from an unknown number, but would be suspicious and extra careful when doing so. Finally, one \textit{legally blind} and one \textit{sighted} participant stated that they set up their phone such that it would not ring for a number not in their contact list.

As previously mentioned, the area code was a secondary indicator participants referred to upon receiving an unsolicited call (17 \textit{legally blind} and 17 \textit{sighted}). For unknown phone numbers, participants would not pick up if they did not recognize the area code or the call was not a local one (four \textit{legally blind}, five \textit{sighted}). Five \textit{legally blind} and two \textit{sighted} participants explained they would evaluate the legitimacy of a call based on people they know or calls they expect from a certain area. Two \textit{legally blind} participants expressed that this indicator became less useful after they moved out of state while retaining their phone number as local calls would no longer match their own area code.

Regarding the STIR/SHAKEN indicator, four \textit{legally blind} and 12 \textit{sighted} participants indicated that they never pick up calls that carry a ``scam likely'' designation. Overall, 24 \textit{legally blind} and 20 \textit{sighted} participants used such a label as part of their decision process when receiving an unsolicited call. The high number of \textit{legally blind} users utilizing this feature suggests that it is well accessible using screen readers. However, two \textit{legally blind} participants also described cases where this method failed, either due to wrongfully labeled benign calls [\textbf{PB5\textsubscript{IS}}] or when an unknown call is expected and therefore accepted before the label can be processed by the screen reader [\textbf{PB13\textsubscript{IB}}]:

\begin{quote}
    ``\textit{Because, honestly, I was not looking for the scam likely tag, because it said blocked number. And I was like, oh, that's my doc, so I didn't even get to the part where the voiceover would have told me. And it does read the whole phone number. Then it says, scam likely, so the idea of an audio (warning) just ahead of the incoming call is very intriguing.}''
\end{quote}

Participants also looked out for scam cues during a call in progress. One indicator they used was whether a recorded or artificial voice was used in the call. Not having a real human on the other end of the line was heavily associated with phishing and scam calls (18 \textit{legally blind} and 12 \textit{sighted}). Participants stated that they recognized ``\textit{familiar voices used in such messages from TikTok}'' [\textbf{PS4\textsubscript{IC}}]. Five \textit{legally blind} and one \textit{sighted} participant declared they would immediately hung up when hearing artificial speech. Another cue utilized in this context was a delay at the start of the call, reported by six \textit{legally blind} and seven \textit{sighted} participants and mostly used in combination with the aforementioned synthetic voice. 

When participants got deep enough into a call, they would assess information on the context they gathered. They demonstrated awareness of common scams and reported using that knowledge to detect fraudulent calls (ten \textit{legally blind} and five \textit{sighted}). Nine \textit{legally blind} and five \textit{sighted} participants explained how authorities would not simply contact them by phone without any previous communication by other means. Similarly, nine \textit{legally blind} and two \textit{sighted} participants mentioned that authorities would not ask for their private information on the phone, nor pressure or urge them into doing something. Finally, ten \textit{legally blind} and ten \textit{sighted} participants explained how they would detect a scam call if it was about an unrealistic scenario. Incidents ranged from services a person never claimed, offers \textit{``too good to be true''}, all the way to nearly impossible situations such as \textit{legally blind} people being contacted regarding their car insurance, as \textbf{PB16\textsubscript{IB}} explained:

\begin{quote}
    ``\textit{One that always gets me, because I'm blind, and there would be no need for me to have car insurance, really!}''
\end{quote}

In terms of their usual responses to scam calls, participants reported taking similar actions to those observed in our cold calls: (1) routinely declining calls, (2) sending them to voice mail and inspecting the transcript at a later time, and (3) answering the call and listening to the content before making a decision. During the interviews, participants described one additional action, namely calling back the alleged source of the call. They do so for any authority which number was publicly known (e.g., the IRS) or familiar to them (e.g., their bank). For unknown sources, participants said they would \textit{``google the number to see if it's anything legitimate''} [\textbf{PS65\textsubscript{IC}}]. 

\subsection{RQ3: Usability of Contextual Warnings}
% Q7 and Q8 and Q10 
% What usability perspectives do (a) \textit{legally blind} and (b) \textit{sighted} people have about telephone scam aural warnings?
We asked all of our participants about the usability of the warnings we tested in our study. We shared all of the aural warnings with all of participants, per scenario, so they had the opportunity to hear the variants we used in our study. The preferences, per scenario, per warning, per group are given in Table~\ref{tab:usability}. The majority of the \textit{legally blind} participants preferred both types of \textit{aural} warnings, expectedly. There is a noticeable shift in preferences towards the contextual warning in the \textit{Social Security} scenario as our participants felt it is particularly useful to give a context to a scam that ``\textit{comes close to heart to blind people whose livelihood largely depends on Social Security benefits}'' [\textbf{PB24\textsubscript{FC}}]

The situation with the \textit{sighted} participants appears quite different. All but one participant in the \textit{interest rate reduction} condition and three in the \textit{Social Security} condition preferred a visual warning akin to the one shown in Figure~\ref{fig:visualwarning}. They didn't dismiss the aural warnings altogether, though, and the majority of them found the context useful, particularly the one warning them about the perils of discussing the Social Security number over the phone. The most prevalent justification for conveying context through a visual alert was the similarity with email or other similar alerts, that, in the words of the participants ``\textit{it's right there in red and grabs your attention right away}'' [\textbf{PS33\textsubscript{FB}}].

When probed to consider the warnings in a broader usability and accessibility context, all participants recalled scenarios where the aural warnings would have an advantage, for example where one could configure ``\textit{an accent of their choice},'' ``\textit{speed},'' and ``\textit{volume}'' of the delivery. They also saw the aural warnings as an excellent way to raise awareness about various scams as they emerge without the need to keep tabs themselves. In the word of \textbf{PS51\textsubscript{IS}}, the contextual aural warning...: 

\begin{quote}
``\textit{...goes along with the adage of like, `give a man a fish he'll eat for a day; teach him to fish, he'll eat for the rest of his life'. So, the contextual warning is far better at educating people about what scams are there and what damage they can do to you.}''    
\end{quote}

We found that the voice of the warning being used in the warning was perceived as a confounding factor in the protection against scam. Some stated that the masculine voice we used is too similar to those employed in scams, and as such, it might startle some people that are not yet used to it. Other participants, however, declared that it's good to grab the attention of the listener, is clear, loud and \textit{``gets the message across''} [\textbf{PB25\textsubscript{IC}}]. Also some participants attested that the voices we used as appropriately authoritative. The female voice was described as calmer, which some preferred. It was perceived as \textit{``trying to educate or warn you''} [\textbf{PB10\textsubscript{FC}}], whereas the male one was more related to solving a problem. One participant suggested recording their warnings in their own voice as a security mechanisms, since scammers could not easily spoof that. Overall the voice was labeled \textit{``super important''} and should convey \textit{``the idea of trust and rapport''}. Especially for blind people, voice is crucial in authenticating the person or organization at the other end of the line, as further explained by \textbf{PB7\textsubscript{FB}}:

\begin{quote}
``\textit{As a blind person, you associate the voice with the brand, in perhaps the same way a sighted person would recognize most of the Chase bank branches look similar.}''
\end{quote}

\begin{table}[!h]
\renewcommand{\arraystretch}{1.1}
% \footnotesize
% \renewcommand{\tabcolsep}{2mm}
\centering
\small
\aboverulesep=0ex % Solution part 1 of 3
   \belowrulesep=0ex % Solution part 1 of 3

\caption{Scam Warnings -- Usability Preferences}
\label{tab:usability}
\begin{tabularx}{\linewidth}{|rYY|}
\toprule

\textbf{Choice} & \textbf{Interest Rate} & \textbf{Social Security} \\\hline

\multicolumn{3}{|c|}{\textbf{Legally Blind Participants}} \\\hline

\textbf{Short warning} & 4 & 0\\
\textbf{Contextual warning} & 0 & 8 \\
\textbf{Both warnings} & 14 & 10\\\hline

\multicolumn{3}{|c|}{\textbf{Sighted Participants}} \\\hline

\textbf{Short warning} & 2 & 2 \\
\textbf{Contextual warning} & 8 & 11 \\
\textbf{Both warnings} & 8 &  5\\\hline
\textbf{Visual and aural} & 1 & 3 \\\hline

\end{tabularx}
\end{table}

\subsection{RQ4: Privacy Concerns and Design Input}
% Q9 and Q11
\subsubsection{Adoption of Contextual Scam Warnings}
Given the intrusiveness of the ``screening'' required to generate a contextual telephone scam warning, we asked all of our participants about whether they would opt-in (or out) for such a service and what privacy concerns they might have. The breakdown, per group, per scenario, and per action/privacy concerns is shown in Table~\ref{tab:privacy-concerns}. Six of the \textit{legally blind} participants in the \textit{interest rate scenario} were reluctant to use these warnings. One of them, a lawyer, said that ``\textit{it's neither legal nor preferred for one to listen to protected calls with clients}'' [\textbf{PB18\textsubscript{IB}}]. The others were hesitant because they invoked negative experiences with ``\textit{targeted Facebook ads that clearly listen to [their] private conversations through the phone's microphone in real-time}'' [\textbf{PB3\textsubscript{IS}}] and lack of transparency about ``\textit{who, how, and where, someone is going to use their data to train yet another AI engine}'' [\textbf{PB25\textsubscript{FB}}].

\begin{table}[!h]
\renewcommand{\arraystretch}{1.1}
% \footnotesize
% \renewcommand{\tabcolsep}{2mm}
\centering
\small
\aboverulesep=0ex % Solution part 1 of 3
   \belowrulesep=0ex % Solution part 1 of 3

\caption{Privacy Concerns - Contextual Warnings}
\label{tab:privacy-concerns}
\begin{tabularx}{\linewidth}{|rYY|}
\toprule

\textbf{Choice} & \textbf{Interest Rate} & \textbf{Social Security} \\\hline

\multicolumn{3}{|c|}{\textbf{Legally Blind Participants}} \\\hline

\textbf{Opt-out (Yes)} & 6 & 6 \\
\textbf{Opt-in (Yes)} & 9 & 7\\
\textbf{Opt-in (No)} & 3 & 5 \\\hline

\multicolumn{3}{|c|}{\textbf{Sighted Participants}} \\\hline

\textbf{Opt-out (Yes)} & 5 & 4 \\
\textbf{Opt-in (Yes)} & 7 & 6 \\
\textbf{Opt-in (No)} & 4 & 6 \\\hline

\end{tabularx}
\end{table}

The majority of the \textit{legally blind} participants in the \textit{interest rate reduction} scenario were fine with opting in, though they expressed concerns about potential privacy infringements. The main concern was the misuse of the terms of use and ``\textit{sending the data back to Google for `improving the service'}'' [\textbf{PB15\textsubscript{IC}}] as participants were open to giving feedback when the warning was wrongly assigned to a given call. \textbf{PB21\textsubscript{IB}} mentioned that this ``\textit{will be just another part of [their] data to be collected}'' as Google already provisions their Gmail, internet, and voice, and felt that this adds to the ``\textit{sense of entirely losing control over who tracks [their] data and for what purpose}.'' Those that had no privacy concerns were open to opt-in because ``\textit{[the warning] does help with just saving time screening calls}'' [\textbf{PB6\textsubscript{IC}}].

In the \textit{Social Security} scenario, three \textit{legally blind} participants stated they would refrain form this client-site solution as they didn't like the idea of someone else ``\textit{listening to their calls, even if its for just few milliseconds}'' [\textbf{PB20\textsubscript{FS}}] or were afraid their ``\textit{data would end up on the dark web somewhere}'' [\textbf{PB4\textsubscript{IB}}] Those that were okay to opt-in were careful to point out that they would do so only on the condition that the data is not transferred outside of their device and have assurances that it would be ``\textit{deleted no later than 15 or 30 days}'' [\textbf{PB33\textsubscript{FC}}] Those that welcomed the production of contextual warnings on the expense of sampling their conversations thought that this approach would, on a long run, help people ``\textit{stop getting these dumb calls for sure}'' [\textbf{PB29\textsubscript{FB}}]. [\textbf{PB42\textsubscript{FC}}] appreciated the warnings as they ``\textit{felt more natural to them as a voice interrupts an unsolicited call and in a friendly manner basically asks you: Are you sure you want to do this?}''

All of the \textit{sighted} participants in the \textit{interest rate reduction} scenario were hesitant about the solution itself (not the warnings, per se) because they had a ``\textit{fraught sense of what AI is doing and how quickly it's taking a presence in a lot of spaces}'' [\textbf{PS61\textsubscript{IC}}]
Those that were open to opt-in had similar reservations expressed as ``\textit{it felt weird just knowing that there's an AI listening all the time}'' [\textbf{PS57\textsubscript{IS}}] Those that were open to opt-out without any privacy concerns reasoned that ``\textit{[their] carriers anyhow are selling [their] data to third parties that contribute to increase in scam calls so people have only their devices to protect themselves from incontinent scam calls}'' [\textbf{PS72\textsubscript{IC}}].

The four \textit{sighted} participants in the \textit{Social Security} scenario that weren't open for the client-based solution, again, resorted to negative experiences with other privacy-related issues, such as cookie consents. \textbf{PB68\textsubscript{FS}} stated that they ``\textit{resent the abundance with which our system is driven by constant marketing and advertising, and really try to keep that noise to a minimum}'' so they would decline any such an intrusion, even if it offered an obvious benefit. Those that were okay to opt-in but had reservations were mostly concerned about the false positives of the approach and the lack of transparency ``\textit{how AI would compensate for them on the expense of private information}'' [\textbf{PB51\textsubscript{FB}}] The participant that welcomed the solution as-is, felt that it's much better option than the STIR/SHAKEN \texttt{Scam Likely}'' because the contextual warning helps``\textit{for the sake of safety and also not wasting time}''
[\textbf{PB54\textsubscript{FC}}]. 

\subsubsection{Design of Contextual Scam Warnings}
As the contextual warning, in both visual and aural variants, are still in an experimental phase (or at least, not mass rolled out as part of mobile OSs), we asked all of our participants to share ideas on how they would prefer being warned, cued, or alerted about incoming scam call ``on the spot.'' An interesting theme emerged among both \textit{legally blind} and \textit{sighted} participants regarding the population that would benefit the most from the \textit{aural} warnings in both the short and contextual variants. Participants felt that older adults would particularly benefit from contextual aural warnings because, as one \textbf{PB14\textsubscript{IS}} put it: ``\textit{they don't know all the scams, so the spoken context without the need to do anything but just pick up the phone, would certainly help them}.''

The \textit{legally blind} part of our sample enjoyed the warnings and liked both the short and the contextual one, predicated that the text remains as the one we came up with at the end of the pilot study. \textbf{PB11\textsubscript{IS}} commented that it would be good for blind or low vision individuals ''\textit{to be able to select a language for the aural warning, for example, Spanish}'' as there are many legally blind people in the US identifying as Hispanic. The blind/low vision participants also liked the alteration of the feminine/masculine tones between the warning and the scam text, but they felt an actual delimitation ``\textit{with one tone or two short consecutive tones}'' [\textbf{PB3\textsubscript{IC}}] would be more usable as that ``\textit{prepares them to approach the scam message itself more alerted}.'' Haptics were recommended frequently as an avenue of alerting people with not just visual disabilities but also deaf and hard of hearing or deafBlind people. Having a ``\textit{default vibration pattern}'' for regular calls and a ``\textit{preceding, different short vibration}'' [\textbf{PB15\textsubscript{IC}}] before the \textit{aural} warning starts would beneficial for multiple participants, especially in busy noisy environments like public transit.  

The \textit{sighted} part of our sample, in equal capacity, enjoyed the aural warnings though all of them preferred the visual one. Many of them mentioned this preference based on the ``\textit{convenience of not having to listen to the conversation}'' [\textbf{PS39\textsubscript{FB}}] and that they ``\textit{just wait for a short second to see if a warning screen will pop up}'' [\textbf{PS46\textsubscript{FS}}] to hang up straight away. But, the majority commented on the ambiguous context about the ``\textit{banks moving money}'' as an action that is confusing and that they prefer to have the aural text variant instead (those that heard it). Few participants commented that the visual warning doesn't resemble the known functions of the green circular button for accept (``dismiss \& continue'') and the red one for hang up (``end call''), so, it is \textit{confusing and not intuitive what action is the best to take here}'' [\textbf{PS45\textsubscript{FB}}]. The (un)familiarity gave a pause to a few participants who wondered whether people ``\textit{would effectively ignore the context and simply press the `end call' button}'' [\textbf{PS50\textsubscript{IB}}] (e.g., by habituation). Here they saw the utility of having an \textit{aural} warning from time to time so people \textit{hear} the context and ``\textit{remain vigilant}'' [\textbf{PS69\textsubscript{FC}}].

% Asl interpretation relay calling with an interpreter on a on a video screen
\section{Discussion}

\subsection{Usability and Accessibility Implications}
Our results suggest a slight shift in the way people might respond to telephone scams in naturalistic settings, compared to the observations by Tu et al.~\cite{Tu2019}. Few years later, people still ``hate'' scam calls~\cite{Tu2016} and rely on the CallerID to screen calls~\cite{Tu2019}, though we saw increased screening by sending the call to ``voice mail.'' This is reasonable to expect as the transcription got better and the voice mail offers a buffer available for screening the call, particularly for legally blind users. We add to the evidence in~\cite{Prasad2020, Prasad2023} that this at-risk population is, in fact, scam calls' targets and they do welcome the idea of contextual warnings. 

To this, our results confirm the findings in~\cite{Edwards2020} that the STIR/SHAKEN indicator (e.g., \texttt{Scam Likely}) does help fend off at least some of the unwanted calls, even if spoofing of CallerID has never subsided at all~\cite{Prasad2023, ftc-data}. The aural warnings we tested overcame the limitation of not having the context of the call identified by Edwards et al.~\cite{Edwards2020}. Compared to the virtual assistant proposed in~\cite{Pandit2023}, the aural warnings eliminate the need for an intermediary and introduce a delay while retaining the agency of the receivers to ultimately decide how to proceed with the call. This is critically important for legally blind users as we found that the STIR/SHAKEN indicator (e.g., \texttt{Scam Likely}) creates a problem where the screen readers don't verbalize the actual CallerID in cases when the call is wrongfully labeled, leaving them without the opportunity to accept a legitimate call. Of course, the contextual warnings are predicated on the screening of the actual call, but privacy concerns also remain in the case of the virtual assistant in~\cite{Pandit2023}. 

Compared to the solution proposed by Du et al.~\cite{Du2023} our results suggest that the contextual warnings, in the aural variant, could coexist with any call-blocking policies, especially for updating the list of predefined call contexts as the scams evolve over time. The same goes for any ``anti-robocall'' apps as they already implement some of the visual elements of the contextual warning shown in Figure~\ref{fig:visualwarning}, per the evaluation done by Sherman et al.~\cite{sherman2020you}. We confirm the findings in~\cite{Sherman2020a11y} relative to the plain language of scam warnings demanded by blind or low vision users, and, our results show that added verbosity (context) imposes no accessibility barriers too. The wide acceptance of both the contextual and the short warning in our study suggests that they could also be complementary augmented with the idea of communicating the accuracy of the various telephone scam call warnings as suggested by Munyaka et al.~\cite{Munyaka2024} (though using a specific alter tone instead of signage).

% The confusion was also a reason for Voight et al. to go beyond just the warning within the incoming call and propose a safe call answering solution for older adults that involved an NFC card~\cite{Voigt2023}. The idea is to build up a trusted source of callers to enable an accompanying card to give a warning to incoming calls from unfamiliar callers.

% The contextual cue about the nature of the scam thus comes closely to the idea of just-in-time and just-in-place warnings implemented in the case of (scam) emails and SMS texts~\cite{Volkamer2017, Petelka2019}.

\subsection{Privacy Implications} \label{sec:privacy-discussion}
The implementation of the aural warnings tested in our study is predicated on scanning the users' conversational data to infer the context in the first place~\cite{GoogleScamDetection}. Done through an AI engine, obviously, this approach raises alarms about the peril of privacy intrusions as well as censorship. In addition to offering the explicit option for users to \textit{opt-in}, we believe that Google or other device manufacturers should strongly work on the \textit{visibility} of such solutions. To this goal, we see the recommendations provided by Chan et al.~\cite{Chan2024} as a good starting point. People using contextual warnings must know the \textit{identity of the AI agent} at any point of time, especially when using the aural variant because there is an AI agent that analyzes the call and another AI agent that ``voices'' the warning itself. Equally, people should have the ability to \textit{monitor in real-time} the scam prevention to flag and potentially filter our problematic behavior. This is especially important in cases of false call labeling that particularly affect the legally blind population. Lastly, the option could exist for FTC to potentially obtain \textit{activity logs} to investigate reports of both emergent scams as well as overly intrusive handling of such calls.

% Activity logs h

\subsection{Limitations}
The naturalistic settings impose several limitations pertaining to our study. A limitation comes from the sample size, the telephone service providers used by our participants, and the top scams per the FTC~\cite{ftc-scams} during 2024.   We were limited to Google's concept of contextual warnings -- future concepts or concepts from other device vendors, notably Apple, might yield results that differ from ours (equally, concepts imposed by regulators in the US). Another limitation is that we tested the warnings in the native English variant to known US scams. Other language implementations of the warnings might cause both legally blind and sighted people to respond in ways that differ from the ones we captured in our study. 

A limitation comes from the fact that we did not test the AI-based generation of the warning as that would have necessitated the use of the experimental Android version offered by Google. As our study was situated in naturalistic settings, we didn't want to confine to only one device and we wanted to study both of our groups using their \textit{own} devices. Besides, the risk of privacy intrusion in our research settings would have been very high and might not justify the benefits of the study. A limitation comes from the synthetic voices we used to generate the warnings and other ``voices'' might produce a different alerting effect than ours. 

% The current version of assistive technologies our participants used as part of the study could pose yet another limitation as any new predictive features (e.g., advanced ``smart glance'') might transform how blind or low vision users access visual warnings on their home screens, and with that, affect the overall findings.

We didn't actually scam our participants for their bank accounts or their Social Security numbers (even the last four digits, as in~\cite{Tu2019}). We only went as far as the first step in the scam -- the ``press one'' aspect -- but we cannot say with certainty whether the warnings we tested would help prevent people, legally blind or sighted, from giving up their details as the scam logic progresses. This was a necessary compromise to balance participants' privacy and well-being while introducing and testing a novel, more robust way of alerting people ``on the spot'' and on their \textit{own} phone number (away from laboratory or settings not naturally occurring for participants, especially the legally blind). The naturalistic settings imposed a limitation to the degree to which the STIR/SHAKEN indicators were implemented; any future updates to this protocol and the way calls are authenticated would certainly affect how people behave around scam calls, and with that, around contextual warnings. We were also limited to the choice of an innocuous phone number as the CallerID. Another CallerID spoofing a known area code or a known number, as our results indicate, might affect how our participants are cued relative to a potential scam call. 

% Similarly, a limitation comes from the choice of spam/junk emails our participants arbitrarily selected in our study as any other message or an encounter with a legitimate email marked as spam/junk might have caused a different behavior around the banner warnings. Though we left our participants sufficient time and support to engage with the emails and the email banner warnings through the assistive technology of their choice, this might have not been insufficient for them to formulate a more informed expression about their overall suspiciousness assessment and ultimate decision about it. Despite all these limitations, and similar to qualitative studies in general, our study nonetheless provides rich accounts of BLV individual's' lived experiences with suspicious emails, which studies in laboratory settings hardly offer. 

\section{Conclusion}
Communicating the context of an incoming scam call does help fend off unwanted calls, regardless of the receiver's visual (dis)ability. In all twelve study conditions involving either six legally blind or six sighted individuals, we found that the \textit{aural} way of altering the caller's intent, just-in-time and just-in-place, is the preferred cue against scams, even in the presence of CallerID indicators or voice mail screening. The implementation of these warnings is predicated on ensuring people's conversational data privacy and we share the cautiously optimistic sentiment expressed by all the participants towards usably accessible solutions against telephone scams of any sort.

\bibliographystyle{IEEEtran}
\bibliography{references}

% \newpage
% \appendix
\appendices

\section{Recruitment Email} \label{sec:recruit}

\noindent \textbf{From}: Researcher's Email \\
\textbf{Subject}: Research Study Participation \\
% \textbf{Date}: \\
\textbf{To}: Potential Participant

\vspace{0.5em}

\noindent Hello, 

\vspace{0.5em}

\noindent My name is \censor{Filipo Sharevski, Ph.D.}, a professor at \censor{DePaul University's School of Computing and Digital Media}. I am contacting you about a possibility to participate in a research interview
discussing general experiences with unsolicited online communication. If you agree to participate, you will be asked to join us remotely via Zoom or phone for a 45-minute interview where we will ask a list of predefined questions.  \\

\noindent If you participate in our research study will be compensated with a \$20 Amazon gift card which will be sent via the same email (the email will notify you that you have \$20 gift card that you can redeem on Amazon) you have provided for contact, following the research interview. \\

\noindent If you are interested in participating, please use the following Calendly link to find and schedule a timeslot that best suits your availability. Once you choose a timeslot and confirm in Calendly, you will receive an email including details on how to join the research interview which will be hosted via Zoom or a phone call. \\

\noindent \censor{https://calendly.com/filipo-sharevski/deceptive-voice-content} \\

\noindent If you have any questions, concerns, or are unable to use the above Calendly link to schedule your interview, please feel free to reach out directly to me either via email, text or phone call. You can also respond directly to this email. My email is \censor{fsharevs@depaul.edu} and my cell phone number is \censor{+1 765-714-9574}. \\

\noindent I thank you for your time in reading this email. \\

\noindent Warm regards, \\
\censor{Filipo Sharevski, Ph.D.}

\section{Interview Script} \label{sec:script}

\subsection*{Announcement}
This interview is being audio-recorded for research purposes. You may stop the recording at any time. Do you consent to being audio-recorded? Recording starts now.

\subsection*{Questions and Tasks}
\begin{enumerate}
\itemsep 0.5em

\item You might have received a phone call in the past that included an unsolicited request in relation to your Social Security number, tax return, utility bill, online accounts, or bank accounts. Tell us more about how you experience and handle such voice calls? 

    \begin{enumerate}
        \item [1.1] Have you noticed anything unusual about these phone calls? Please specify in as much details you can. 

        \item [1.2] Have you noticed any warnings, notifications, or labels about these calls (e.g. ``Scam Likely'' as a callerID)? Please specify in as much details you can. 

        \item [1.3] How do you usually review and decide what to do with these calls?

        \item [1.5] What cues do you usually use to assess the legitimacy of a phone call?

        \item [1.6] How often you receive potentially deceptive voice calls?

        \item [1.7] Have you ever used any phone manufacturer (e.g., Apple, Google, Samsung, etc.) or cellphone carrier (e.g., T-mobile, ATT, Verizon, etc.) feature to block, restrict, or screen deceptive voice calls? 
        
    \end{enumerate}

\item You might have received a phone call in the past that contained a message warning you that the call \textit{might be potential scam}, containing an unsolicited request in relation to your Social Security number, tax return, utility bill, online accounts, or bank accounts. Tell us more about how you experienced and handled this voice call? 

    \begin{enumerate}
        \item [2.1] What was your experience with this voice warning? Please specify in as much details you can. 

        \item [2.2] Have you noticed any other warnings, notifications, or labels about these calls (e.g. ``Scam Likely'' as a callerID) in addition to the voice warning? Please specify in as much details you can. 

        \item [2.3] Would this voice warning factor in the way you review and decide what to do with these calls? If so, could you provide us more details how? If not, could you tell us why not? 

        \item [2.4] What you be interested to use such a feature to warn you about the nature of a phone call before you connect to the call?

        \item [2.5] These warnings are custom created by us as researchers, but for them to be implemented in practice, the phone manufacturer (e.g., Apple, Google, Samsung, etc.) or cellphone carrier (e.g., T-mobile, ATT, Verizon, etc.) might need to listen for conversation patterns commonly associated with deceptive calls to be able to use an AI and apply the warning in a dynamic fashion. How would you feel about such a practice?  
        
    \end{enumerate}

\item What is your perspective on how your cellphone carriers handle deceptive phone calls? 

\item Have you ever been a victim of a successful deceptive phone call(s)? If you are comfortable with, please share your experiences with this event(s). [What lessons you have learn from here and how this episode affected your way of dealing with potentially deceptive phone calls afterwards]

\item Have you seen any other types of deceptive campaigns delivered through voice message but over other types of communication than phones (e.g. WhatsApp, social media, etc.)?

\item What would you recommend about how these warnings be designed adequately accessible for blind people or people with visual impairments?

\item Anything else you want to add on this topic or your experience with warnings about deceptive voice calls?

\item Demographic Questions

\begin{enumerate}
\item How old are you?
\item What race/ethnicity do you identify as?
\item What is your gender?
\item What is your visual diagnosis?
\item What is your education level?
\item What is your cellphone provider?
\item What is your experience with deceptive voice calls? 
\end{enumerate}

\end{enumerate}

\section{Codebook} \label{sec:codebook}

% DATASET: https://docs.google.com/spreadsheets/d/1UoHSDT_Uyfk1l3y7EkGmZrf1aU1_DthB/edit?usp=drive_link&ouid=106132403222903103013&rtpof=true&sd=true

\subsection{Initial Response} 
\noindent Codes pertaining to the initial response to the study phone calls, some of which included contextual scam warnings.

\begin{enumerate}
\itemsep 0.5em 

% These are based on analysis of CALLFIRE LOGS
\item \textbf{Action} Codes pertaining to the \textit{action} taken by the participant when they received the study phone call.
        
        \begin{itemize}
            \item \textbf{Answered, Pressed 1} The participant expressed that they answered the phone and pressed 1
            
            \item \textbf{Answered, hang up} The participant expressed that they answered the phone but hung up after listening to the content
            
            \item \textbf{Let it go to voice mail} The participant expressed that they let the phone call go to voice mail and then access the recording/transcript
            
            \item \textbf{Declined} The participant expressed that they declined the incoming voice call
        \end{itemize}

% Q1. Did you receive any calls unsolicited calls in the past few days?
\item \textbf{Recall} Codes pertaining to the \textit{recall} by the participant about  recent experience with an unsolicited call.

        \begin{itemize}
            \item \textbf{Yes, confirming the study call} The participant expressed that they received a call and the call was the one initiated by the researchers (participants were debriefed at this point and consent for retaining their data was obtained)

            \item \textbf{Yes, but not the call from the study} The participant expressed that they received a call but the call was not the one initiated by the researchers 
            
            \item \textbf{No} The participant expressed that they never received any unsolicited call
            
        \end{itemize}

% Q2. What did you notice about this phone call?
\item \textbf{Cues} Codes pertaining to the \textit{cues} noticed by the participant about their recent experience with an unsolicited call, if they had any.

        \begin{itemize}
            \item \textbf{Unknown Number} The participant expressed that they noticed that call was from an unknown number
            
            \item \textbf{Area Code} The participant expressed that they noticed that the area code differs from the area code of their phone

            \item \textbf{CallerID: Spam or Scam Likely} The participant expressed that they noticed that CallerID was labeled as Spam or Scam Likely

            \item \textbf{Synthetic Voice} The participant expressed that they noticed that voice of the caller sounded synthetic

            \item \textbf{Delay} The participant expressed that they noticed that there was a delay between answering the call and the speech on the other side

            \item \textbf{Warning} [Except the baseline groups] The participant expressed that they noticed that a warning was communicated to them about the call

        \end{itemize}

\end{enumerate}

\subsection{Reflection} 
\noindent Codes pertaining to the response upon reflection about the study phone calls, some of which included contextual scam warnings.

% Q3. Would you have picked it up? 

\begin{enumerate}
\itemsep 0.5em
\item \textbf{Action upon Reflection} Codes pertaining to the \textit{action} that participant might have taken upon the reflection on the cues relative to the study (or other unsolicited) phone call(s).
        
        \begin{itemize}
            \item \textbf{Answered} The participant expressed that they would have answered the phone and pressed 1
            
            \item \textbf{Answered, listened to the content} The participant expressed that they would have answered the phone but hung up after listening to the content
            
            \item \textbf{Screened the voice mail} The participant expressed that they let the phone call go to voice mail and then access the recording/transcript to screen the call 
            
            \item \textbf{Declined} The participant expressed that they would have declined any incoming voice call they don't recognize 
            
        \end{itemize}

% Q4. Have you gotten calls like this before?
\item \textbf{Experience with Unsolicited Calls} Codes pertaining to the \textit{experience (e.g. the frequency)} with unsolicited calls in general.

        \begin{itemize}
            \item \textbf{No experience} The participant expressed that they haven't received any call 

            \item \textbf{Never answered them} The participant expressed that they have received unsolicited calls but they never answer them 
            
            \item \textbf{Yes, seldom} The participant expressed that they seldom receive unsolicited calls
            
            \item \textbf{Yes, frequently} The participant expressed that they frequently receive unsolicited calls
            
            \item \textbf{Yes, all the time} The participant expressed that receive unsolicited calls all the time (once or multiple a day, every day) 
            
        \end{itemize}

% Q5. How do you usually asses if a call is spam?
\item \textbf{Dealing with Unsolicited Calls} Codes pertaining to the way participants \textit{deal} with unsolicited calls in general.

           \begin{itemize}
            \item \textbf{Unknown Number} The participant expressed that they are suspicious when a call comes from an unknown number
            
            \item \textbf{Area Code} The participant expressed that they are suspicious when the area code of the incoming call differs from the area code of their number

            \item \textbf{CallerID: Spam or Scam Likely} The participant expressed that they are suspicious about a call if they notice that CallerID was labeled as Spam or Scam Likely

            \item \textbf{Recorded Voice} The participant expressed that they are suspicious when they noticed the voice on the line is recorded 
            
            \item \textbf{Synthetic Voice} The participant expressed that they are suspicious when they noticed the voice on the line is synthetically generated

            \item \textbf{Delay} The participant expressed that they noticed that there was a delay between answering the call and the speech on the other side

            \item \textbf{Context}  The participant expressed that they are suspicious when the call is about personal information or about an inapplicable scenario (a car insurance expiration for blind people, for example)

        \end{itemize}

\end{enumerate}

\subsection{Contextual Telephone Scam Warnings} 
\noindent Codes pertaining to the reception of the scam warnings used in the study (note, for those participants in the baseline condition, we played both recordings during the interview, after debriefing them about the goal of the study)

% Q6. What is your opinion about this warning? Considered together with
%  Q7. Would you be interested in a long one that describes not just the warning?
\begin{enumerate}
\itemsep 0.5em
    \item \textbf{Usefulness} Codes pertaining to the \textit{usefulness} of the warning in fending off scam calls 

        \begin{itemize}
            
            \item \textbf{Short warning useful} The participant expressed that they found the short warning useful 

            \item \textbf{Long warning useful} The participant expressed that they found the short warning useful 

            \item \textbf{Both short and long warning useful} The participant expressed that they found both the short and the long warning useful 

            \item \textbf{Short warning useful; long warning not} The participant expressed that they found the short warning useful, but the long one not 

            \item \textbf{Long warning useful; short warning not} The participant expressed that they found the long warning useful, but the short one not
            
        \end{itemize}
    
% Q8. Recommendations  

    \item \textbf{Population} Codes pertaining to the \textit{target populations} the would benefit the most from the contextual telephone scam warnings  

        \begin{itemize}
            
            \item \textbf{Older People} The participant recommended using the contextual telephone scam warnings for alerting older people

            \item \textbf{Blind/Low Vision People} The participant recommended using the contextual telephone scam warnings for alerting blind/low vision people

        \end{itemize}

    \item \textbf{Design} Codes pertaining to recommended \textit{designs} contextual telephone scam warnings   

        \begin{itemize}

         \item \textbf{Tones} The participant recommended using various \textit{tones} in addition to the warning to alert people with visual disabilities

        \item \textbf{Haptics} The participant recommended using various \textit{haptic patterns} in addition to the warning to alert people with visual disabilities

        \end{itemize}

    \item \textbf{Preference} Codes pertaining to the \textit{preference} of begin alerted about incoming scam call 

            \begin{itemize}
            
            \item \textbf{No warning} The participant expressed that they don't want to be warned about potentially incoming scam call

            \item \textbf{Short Warning} The participant expressed that they prefer the short contextual telephone scam warning

            \item \textbf{Long Warning} The participant expressed that they prefer the long contextual telephone scam warning

            \item \textbf{Visual Warning} The participant expressed that they prefer a visual contextual telephone scam warning

        \end{itemize}

% Q9. Would you be interested to opt into receiving contextual warnings?

 \item \textbf{Use and Privacy Concerns} Codes pertaining to the \textit{use and privacy concerns} about the implementation of contextual telephone scam warnings

      \begin{itemize}
     
        \item \textbf{Opt-in, no privacy concerns} The participant expressed that they would opt-in for contextual telephone scam warnings and have no concerns about any potential intrusions of their privacy

        \item \textbf{Opt-in, but privacy concerns} The participant expressed that they would opt-in for contextual telephone scam warnings but expressed concerns about any potential intrusions of their privacy

        \item \textbf{Opt-out, privacy concerns} The participant expressed that they would opt-out from using contextual telephone scam warnings and due to, in part, privacy concerns about any potential intrusions of their privacy

      \end{itemize}

 % \item \textbf{Experience with Telephone Scams} Codes pertaining to the \textit{past experiences} with telephone scam warnings

 %      \begin{itemize}
     
 %        \item \textbf{Fell victim} The participant expressed that they have fallen vitim to a telephone scam in the past

 %        \item \textbf{A person they know fell victim} The participant expressed that they a person they know have fallen victim to a telephone scam in the past

 %        \item \textbf{Never fell victim} The participant expressed that they have never fallen victim to a telephone scam in the past

 %      \end{itemize}

\end{enumerate}

\section{Debriefing} \label{sec:debrief}
Thank you for participating in our research on how individuals who are low vision or blind experience telephone scam calls compared to individuals who are sighted. This study aimed to examine how individuals would react to warnings about a potential scam call. The call you received yesterday was initiated by us, the researchers. If you heard any warnings, these were also created by us and the scam message (if you were in the baseline group) that was included too, based on the most popular scams tracked by the FTC. The entire call was innocuous -- any digit that was requested to be pressed was automatically transferred back to a phone number controlled by us the researchers where you had the ability either to speak to them or leave a voice message. So far, no research exists on how low vision or blind individuals, compared to sighted individuals, experience and potentially might utilize telephone scam contextual warnings in naturalistic settings, that is, without knowing \textit{a priori} that they might receive such a  call. This is why we asked you to sign up to the study with the phone and we placed a call 24 hours before the time slot you have selected for participation. \\

\noindent It was necessary for the researchers to withhold this information from you regarding the purpose of the study to ensure that your actions and answers to questions accurately reflected your response and utilization of the short and contextual warnings we created. Your participation in the study is important in helping researchers identify the best ways to address the design, accessibility, and preferential formatting of the warnings assigned by phone providers or phone manufacturers before connecting you to a potentially scam call. You have the option to provide \textit{a posteriori} consent so we retain your action and response from the call and continue with the interview. After we end this meeting and compensation is provided, then we cannot remove your information since we did not collect any personal information (we removed your phone number from our data records). \\

\noindent The final results of this study will be published in a peer-reviewed journal or conference. Your results will not be available individually and your participation will remain confidential. We do not keep, record, or collect any personal information such as telephone numbers. If you have any additional inquiries please contact \censor{Filipo Sharevski, Ph.D. at fsharevs@cdm.depaul.edu}. If you have questions about your rights as a research subject, you may contact \censor{Jessica Bloom} in the Office of Research Services at \censor{(312) 362-6168 or via email at jbloom8@depaul.edu}. You may also contact \censor{DePaul’s} Office of Research Services if your questions, concerns, or complaints are not being answered by the research team, you cannot reach them, or you want to talk to someone besides them. \\

\newpage % The Meta-Review should at least start on a new column

\end{document}